\documentclass[11pt,a4paper]{article}
\pdfoutput=1
\usepackage{ifpdf}
\usepackage{jcappub}
\usepackage{rotating}
\usepackage[bb=boondox]{mathalfa}
\usepackage{comment}

\usepackage{tikz-cd}

\usepackage{graphicx} 
\usepackage{mathrsfs}
\usepackage{amsmath,amsfonts,amssymb}
\usepackage{mathtools}
 \usepackage{booktabs}
\usepackage{graphicx}
\usepackage{dcolumn}
\usepackage{bm}
\usepackage{hyperref}
\usepackage[mathlines]{lineno}
\usepackage{float}
\usepackage{blindtext}
\usepackage{titlesec}
\title{Sections and Chapters}
\usepackage[toc,page]{appendix}
\usepackage{xcolor}
\usepackage{graphicx}
\graphicspath{ {./images/} }
\usepackage[thinc]{esdiff}
\usepackage[euler]{textgreek}
\usepackage{braket}
\usepackage{physics}
\usepackage{xfrac}
\usepackage{soul}
\usepackage{stmaryrd}
\usepackage{pifont}
\usepackage{caption}
\usepackage{makecell}
\usepackage{orcidlink}
\usepackage{soul}
\usepackage{multirow}

\newcommand{\dis}[1]{\begin{equation}\begin{split}#1\end{split}\end{equation}}
\usepackage{slashed}

\usepackage[commandnameprefix=always]{changes}

\allowdisplaybreaks

\title{Leptogenesis and Planck-scale black hole remnants in a Pati–Salam cosmology}

\author[a]{Arnab Chaudhuri\,\orcidlink{0000-0002-6784-1360}\,}
\author[b]{\!\!, Erdenebulgan Lkhagvadorj\,\orcidlink{0009-0004-5953-8267}\,}
\author[c]{\!\!, Satyabrata Mahapatra\,\orcidlink{0000-0002-4000-5071}\,}
\affiliation[a]{School of Advanced Sciences,
Vellore Institute of Technology, Vellore, Tamil Nadu 632014, India.}
\affiliation[b]{Department of Physics and Institute of Basic Science, Sungkyunkwan University, 2066 Seobu-ro, Suwon-si, Gyeonggi-do, 16419, Korea.}
\affiliation[c]{School of Physical Sciences, Indian Institute of Technology Goa, Ponda-403401, Goa, India.}
\emailAdd{arnab.chaudhuri@vit.ac.in}
\emailAdd{bulgaa@skku.edu}
\emailAdd{satyabrata@iitgoa.ac.in}

\abstract{
We study leptogenesis and Planck-scale remnant dark matter from
primordial black hole (PBH) evaporation in a minimal Pati--Salam cosmology,
with the gauge symmetry broken before inflation so that magnetic
monopoles are diluted away. A singlet inflaton with a near-inflection
potential enhances the curvature power spectrum on small scales,
producing a narrow black hole population that briefly dominates the
energy density. The Pati--Salam embedding ties the right-handed neutrino
masses to the $SU(2)_R$ breaking scale, and cosmological consistency then
forces both the scalar-sector Yukawa coupling and the heavy neutrino mass
well below that scale. Two regimes emerge, separated by whether the black
holes are hot enough to emit the lightest right-handed neutrino. A PBH of lighter mass drive non-thermal leptogenesis, while heavier ones require a thermally
generated asymmetry and also affect it through entropy dilution before settling to the final value. If quantum
gravitational backreaction halts evaporation at the Planck scale, each PBH leaves a stable remnant whose present abundance scales as the inverse
five-halves power of the initial mass. Suppressing it therefore requires
heavier black holes, but the lighter PBH needed for
non-thermal leptogenesis overproduce such Planck-scale remnant dark matter. Thus Remnant dark matter and non-thermal leptogenesis are mutually
exclusive. However, once the thermal contribution is invoked for leptogenesis, a single initial PBH mass near
$10^6$ g accommodates the observed dark matter abundance, the heavy
neutrino mass required by the baryon asymmetry, and the Pati--Salam
breaking scale together. The scenario predicts a stochastic gravitational
wave background from the Poisson fluctuations of the black hole
distribution, within reach of future experiments,
alongside high-frequency graviton emission constrained by future measurements of the effective number of relativistic species.

}

\begin{document}
\maketitle
\flushbottom


\section{Introduction}\label{sec:intro}

The origin of the baryon asymmetry of the Universe (BAU) and the fundamental nature of dark matter (DM) remain two of the most profound and persistent open problems in modern cosmology and particle physics. Astronomical observations of the cosmic microwave background and the primordial elemental abundances derived from Big Bang Nucleosynthesis (BBN) indicate a precise baryon-to-entropy ratio of $Y_B^{\rm obs} \simeq 8.7 \times 10^{-11}$, \cite{Planck:2018jri,Cyburt:2015mya}. While the Standard Model (SM) of particle physics successfully accommodates both CP violation through the Cabibbo-Kobayashi-Maskawa (CKM) matrix \cite{Sakharov:1967dj} and baryon number violation through non-perturbative electroweak sphaleron transitions, it falls quantitatively short of explaining the observed asymmetry. The electroweak phase transition, given the measured mass of the Higgs boson, is a smooth crossover rather than a first-order phase transition, thereby failing to provide the required departure from thermal equilibrium dictated by the Sakharov conditions \cite{Sakharov:1967dj}, and the inherent CP violation is insufficient by several orders of magnitude \cite{Gavela:1993ts,Gavela:1994dt,Riotto:1999yt}. Furthermore, the Standard Model entirely lacks a viable, non-baryonic cold dark matter candidate to account for the roughly $27\%$ of the Universe's energy budget.

Leptogenesis presents an elegant, minimal, and highly motivated solution to the baryogenesis problem by linking the generation of the baryon asymmetry to the out-of-equilibrium, CP-violating decays of heavy right-handed neutrinos (RHNs) \cite{Fukugita:1986hr, Kuzmin:1985mm, Minkowski:1977sc,Yanagida:1979as,Gell-Mann:1979vob,Mohapatra:1979ia}. In the canonical thermal leptogenesis paradigm, these heavy states are populated by the thermal bath of the early Universe \cite{Buchmuller:2004nz,Giudice:2003jh}. Their subsequent CP-violating decays generate a primary lepton asymmetry, which is partially reprocessed into a net baryon number by baryon-plus-lepton ($B+L$) violating electroweak sphalerons processes
which convert the $B-L$ asymmetry into a baryon
asymmetry before they become ineffective at temperatures of order the
electroweak scale. This framework is deeply connected to the neutrino sector, as it simultaneously explains the extreme smallness of the active light neutrino masses through the type-I seesaw mechanism \cite{Gell-Mann:1979vob,Mohapatra:1979ia}.

To provide a rigorous, symmetry-motivated origin for these right-handed neutrinos, the leptogenesis mechanism can be naturally embedded within the Pati-Salam gauge structure, $SU(4)_C \times SU(2)_L \times SU(2)_R$ \cite{Pati:1974yy,Pati:1973uk}. In the Pati-Salam framework, quarks and leptons are fundamentally unified by treating lepton number as the fourth color. Right-handed neutrinos arise automatically and inevitably as intrinsic components of the $SU(2)_R$ fermion doublets, making their existence a structural prediction of the gauge geometry rather than an ad hoc phenomenological extension of the Standard Model. Furthermore, the Majorana masses of these heavy right-handed neutrinos are directly tied to the $SU(2)_R$ symmetry-breaking scale through the vacuum expectation value of an extended scalar multiplet. This structural feature removes the arbitrariness often present in generic leptogenesis models, explicitly linking the high-energy mass scale of the right handed neutrinos to the grand unified symmetry-breaking scale.

In standard hierarchical thermal leptogenesis, however, the viable parameter space is tightly restricted by the Davidson--Ibarra bound \cite{Davidson:2002qv}. 
Achieving the observed baryon asymmetry within the
standard thermal scenario therefore generally requires a sufficiently
heavy $N_1$ and a correspondingly high temperature for its efficient
production. In a Pati--Salam cosmology, however, such high
temperatures can be problematic if they restore the broken gauge
symmetry after inflation and triggering the overproduction of dangerous topological relics like magnetic monopoles, which would subsequently overclose the Universe.  
These considerations motivate a non-thermal source of RHNs
that does not rely on their efficient production from the thermal
plasma.
An alternative and highly compelling pathway that cleanly bypasses these high-temperature tensions is the non-thermal production of heavy right-handed neutrinos via the Hawking evaporation of primordial black holes.

Primordial black holes can form in the early Universe following the horizon re-entry of enhanced small-scale curvature perturbations~\cite{Carr:2009jm, Carr:2020gox}, which are typically generated during a transient ultra-slow-roll phase of cosmic inflation. Depending on their initial mass fraction, primordial black holes can temporarily dominate the total energy density of the Universe, ushering in an early matter-dominated epoch~\cite{Choi:2023kxo, Masina:2020xhk,Chaudhuri:2023aiv}. As they evaporate through Hawking radiation, they inject massive amounts of entropy into the surrounding plasma \cite{Chaudhuri:2020wjo} and emit massive particles democratically, regardless of their specific Standard Model gauge couplings. This democratic emission provides a genuinely non-thermal source for right-handed neutrinos~\cite{Fujita:2014hha} and hence leptogenesis. This mechanism therefore relaxes the need
for the thermal plasma to populate the heavy RHN sector at
temperatures of order $M_{N_1}$ allowing for successful baryogenesis at significantly lower reheating temperatures~\cite{Baumann:2007yr, Kawasaki:2000en, Hannestad:2004px}. 

In this manuscript, primordial black hole-assisted non-thermal leptogenesis is embedded within a minimal Pati-Salam cosmology. The analysis mandates that the Pati-Salam gauge symmetry is broken prior to the onset of inflation. This pre-inflationary breaking ensures that any dangerous topological defects, such as magnetic monopoles generated during the phase transition, are exponentially diluted by the inflationary expansion and do not disrupt subsequent cosmological evolution. This temporal sequencing imposes strict cosmological consistency conditions on the model: the Pati-Salam symmetry must not be thermally restored at any point in the post-inflationary history, and the initial black hole temperature must be sufficiently high to kinematically permit the unsuppressed emission of the massive right-handed neutrinos. 

A particularly important constraint arises from the interplay between
PBH evaporation and electroweak sphaleron conversion. In the limiting
case in which the baryon asymmetry is generated predominantly by RHNs
emitted from PBHs, the resulting $B-L$ asymmetry must be produced and
made available to the sphaleron processes before they become
inefficient at $T_{\rm EW}\simeq130~{\rm GeV}$. Requiring PBH
evaporation to be completed before sphaleron freeze-out then yields
the upper bound
$M_{\rm in}\lesssim 4.2\times10^{5}~{\rm g}\,,
$
for the initial PBH mass. This bound is strongest for the
PBH-dominated, PBH-only leptogenesis scenario.  When thermally produced
RHNs contribute to the asymmetry, part of the
$B-L$ asymmetry is generated before PBH evaporation and the constraint
on the evaporation time is correspondingly relaxed.

Beyond the generation of the baryon asymmetry, we also investigate the possibility that quantum gravitational backreaction can halt the Hawking evaporation process as the black hole mass approaches the fundamental Planck scale. Theories such as Loop Quantum Gravity and various non-commutative geometries suggest that the event horizon may be replaced by a quantum-corrected structure, preventing the formation of a physical singularity and halting the evaporation. This process leaves behind stable, non-radiating Planck-scale remnants that serve as highly motivated cold dark matter candidates~\cite{MacGibbon:1987my, Carr:2020xqk, Carr:2016drx, Domenech:2023mqk, Trivedi:2025vry, Lehmann:2019zgt, Ong:2024dnr, Sasaki:2025zao}. Because these remnants are incredibly compact, massive compared to elementary particles, and interact exclusively through gravitational forces, they represent an idealized dark matter state that easily evades all conventional direct detection constraints. Their relic abundance is
therefore determined by the number density of PBHs together with the
remnant mass, and can be directly related to the initial PBH mass.

By analytically tracking the relic density of these Planck-scale remnants, we find that there is a critical and inescapable phenomenological tension between the requirements of
leptogenesis and dark matter production.  For the remnants to account for the entirety of the observed dark matter density, the required initial mass of the primordial black holes strictly exceeds the upper mass bound permitted by electroweak sphaleron conversion. 
Requiring the
remnants to account for the entire observed dark matter abundance
gives
$M_{\rm in}\simeq1.1\times10^{6}~{\rm g}\,$.
This value is larger than the upper bound
$M_{\rm in}\lesssim4.2\times10^{5}~{\rm g}$ obtained when the baryon
asymmetry is generated solely by PBH-produced RHNs and must be
processed by electroweak sphalerons before their freeze-out.
Consequently, within this hierarchical non-resonant setup considered
here, PBH remnants cannot simultaneously constitute all of the dark
matter and realize the PBH-only leptogenesis limit. The tension can
instead be alleviated when thermally produced RHNs contribute to the
$B-L$ asymmetry, or when the remnants constitute only a subdominant
fraction of the observed dark matter density. 

Finally, although the stochastic gravitational wave background associated with the pre-inflationary Pati--Salam phase transition is exponentially diluted during inflation, the primordial black hole population can give rise to several distinct gravitational-wave-related signatures. These include scalar-induced gravitational waves sourced by the enhanced curvature perturbations responsible for PBH formation, tensor perturbations associated with the inhomogeneous PBH distribution and its subsequent evolution, and gravitational radiation generated during PBH evaporation through processes such as direct Hawking emission and bremsstrahlun\cite{Choi:2024acs, LISACosmologyWorkingGroup:2023njw,  Domenech:2024kmh}. In this work, we focus specifically on the stochastic background of gravitons produced directly through Hawking evaporation of the PBHs. The resulting ultra-high-frequency gravitational waves provide a complementary probe of the PBH population and of the underlying cosmological history, and may be relevant for future constraints on additional relativistic degrees of freedom through their contribution to the radiation energy density. These signatures span a vast range of frequencies, providing distinct, testable predictions for future gravitational wave observatories such as the Einstein Telescope (ET)~\cite{ET:2019dnz}, Cosmic Explorer (CE)~\cite{Reitze:2019iox}, and next-generation cosmic microwave background surveys~\cite{Planck:2018vyg, CMB-S4:2016ple, CMB-HD:2022bsz} targeting the effective number of relativistic species.

This paper is organized as follows: we summarize the relevant aspects of the PS model and its connection to PBH-assisted leptogenesis in section~\ref{sec:PS}. Moreover, we discuss the PBH formation scenario during inflation in section~\ref{sec:inflation}, while we review the basics of PBH evaporation and the non-resonant leptogenesis scenario in section~\ref{sec:leptogenesis}. In section~\ref{sec:dm}, we further investigate the Planck scale PBH remnants as a dark matter candidate. In section~\ref{sec:gw}, we point out that our scenario can be tested by the existing and future gravitational wave experiments and finally conclude in section~\ref{sec:conc}.

\section{Pati--Salam framework and cosmological consistency conditions}
\label{sec:PS}

We embed the PBH-assisted leptogenesis scenario studied in this work within the Pati--Salam (PS) gauge framework
\begin{equation}
G_{\rm PS} = SU(4)_C \times SU(2)_L \times SU(2)_R,
\end{equation}
originally proposed in Refs.~\cite{Pati:1974yy,Pati:1973uk}. 

In this framework, quarks and leptons are fundamentally unified by treating lepton number as the fourth color. For each generation, the left-handed and right-handed fermion representations transform as
\begin{equation}
F_L \sim (4,2,1),
\qquad
F_R \sim (4,1,2),
\end{equation}
where the numbers denote representations under $SU(4)_C$, $SU(2)_L$, and $SU(2)_R$, respectively. Because the right-handed multiplet $F_R = (u_R, d_R, \nu_R, e_R)$ explicitly contains the right-handed neutrino $\nu_R \equiv N_R$ without additional assumptions, the existence of heavy Majorana right-handed neutrinos is a rigorous structural prediction of the gauge symmetry rather than an ad hoc extension of the Standard Model.
Thus, the PS framework provides a gauge-theoretic origin for the right-handed neutrinos required for the seesaw mechanism and leptogenesis.

We consider the symmetry-breaking chain
\begin{equation} \label{eq:chain}
SU(4)_C \times SU(2)_L \times SU(2)_R
\;\longrightarrow\;
SU(3)_C \times SU(2)_L \times U(1)_Y
\;\longrightarrow\;
SU(3)_C \times U(1)_{\rm em},
\end{equation}
where the first step occurs at a scale $v_R$ associated with the breaking of $SU(2)_R$. 
The first stage involves the breaking of the PS gauge symmetry, including $SU(4)_C\to SU(3)_C\times U(1)_{B-L}$ and $SU(2)_R\to U(1)_R$, with the hypercharge generator $Y = T_{3R} + (B-L)/2$ arising from the appropriate combination of the $SU(2)_R$ and $B-L$ generators. 
The parameter $v_R$ denotes the vacuum expectation value (VEV) of the scalar multiplet responsible for $SU(2)_R$ breaking. This breaking is driven by a scalar multiplet $\Delta_R \sim (10,1,3)$. The Majorana mass term for the right-handed neutrinos arises from a Yukawa interaction of the form
\begin{equation}
\mathcal{L}_Y \supset
y_R\, F_R^T C^{-1} \Delta_R F_R + \text{h.c.},
\end{equation}
When the neutral component acquires a vacuum expectation value $\langle \Delta_R \rangle = v_R$, the Yukawa interaction generates a Majorana mass matrix for the right-handed neutrinos,
\begin{equation}
M_{N} = y_R\, v_R,
\label{eq:MN_PS}
\end{equation}
with $y_R$ is a dimensionless Yukawa coupling {matrix} and Eq.~(\ref{eq:MN_PS}) is understood as a matrix relation, up to an $\mathcal{O}(1)$ normalization factor arising from the triplet convention, which we absorb into the definition of $y_R$. Thus, the PS framework directly correlates the RHN mass scale with the scale of $SU(2)_R$ breaking. In the hierarchical regime relevant for the leptogenesis analysis, we identify the lightest RHN mass as
\begin{equation}
M_{N_1}=y_R^{(1)}v_R,,
\end{equation}
where $y_R^{(1)}$ denotes the corresponding eigenvalue in the basis in which the RHN mass matrix is diagonal. In the numerical analysis below, we therefore regard $y_R^{(1)}$ as a derived quantity once $M_{N_1}$ and $v_R$ are specified.


Due to the underlying left-right discrete symmetry (D-parity) of the Pati--Salam group, the scalar sector naturally contains a left-handed triplet $\Delta_L \sim (10,3,1)$ alongside the right-handed triplet $\Delta_R \sim (10,1,3)$. If $\Delta_L$ acquires an induced vacuum expectation value $v_L$, it yields an additional type-II seesaw contribution to the light-neutrino mass matrix.  We assume that D-parity is broken at a sufficiently high scale such that the left-handed triplet is substantially heavier than the $SU(2)_R$ lb breaking scale. Its induced VEV is then suppressed, schematically as
$
v_L\propto{v_{\rm EW}^2}/{M_{\Delta_L}},
$
up to model-dependent scalar couplings. We therefore work in the type-I seesaw-dominated limit, in which the contribution from $\Delta_L$ to the light-neutrino mass matrix can be neglected. This assumption fixes the neutrino-mass relation used in the leptogenesis analysis below without introducing an additional type-II contribution.


Below the PS-breaking scale, the effective theory consists of the Standard Model gauge group supplemented by heavy singlet neutrinos of mass $M_{N}$ {(the RHNs, now $SU(2)_L$ singlets)}. Electroweak symmetry breaking is realized through a scalar bi-doublet transforming as $(1,2,2)$ under $G_{\rm PS}$, which contains the Standard Model Higgs doublet after $SU(2)_R$ breaking. We assume that one linear combination remains light and is identified with the Standard Model Higgs doublet.  The Dirac neutrino mass matrix takes the form
\begin{equation}
m_D = y_\nu \frac{v_{\rm EW}}{\sqrt{2}},
\end{equation}
where $y_\nu$ is the Dirac Yukawa coupling matrix and $v_{\rm EW}=246$ GeV is the electroweak vacuum expectation value. Light neutrino masses arise through the type-I seesaw mechanism \cite{Minkowski:1977sc,Yanagida:1979as,Gell-Mann:1979vob,Mohapatra:1979ia},
\begin{equation}
m_\nu \simeq - m_D^T M_{N}^{-1} m_D .
\label{eq:seesaw_PS}
\end{equation}
For the single-flavour approximation employed in the Boltzmann analysis, this relation reduces parametrically to \begin{equation}
y_\nu^2\simeq\frac{m_\nu M_{N_1}}{v^2},\,
\qquad
v\equiv\frac{v_{\rm EW}}{\sqrt{2}}\simeq174~{\rm GeV}.
\end{equation}
Taking $m_\nu\simeq0.05~{\rm eV}$ as a representative light-neutrino mass scale gives
\begin{equation}
y_\nu \simeq 1.3\times10^{-2}
\left(\frac{M_{N_1}}{10^{11}~{\rm GeV}}\right)^{1/2}
\left(\frac{m_\nu}{0.05~{\rm eV}}\right)^{1/2} .
\label{eq:ynu_PS}
\end{equation}
This relation is used below to determine the effective Dirac Yukawa coupling entering the RHN decay width and the leptogenesis dynamics.

Throughout this work, we assume that PS symmetry breaking occurs prior to the onset of inflation. Any monopoles or other topological relics produced during the PS phase transition are therefore diluted away by inflation and do not affect the subsequent cosmological evolution. Consistency of this assumption requires that the inflationary Hubble scale to remain below the characteristic PS-breaking scale
\begin{equation}
H_{\rm inf} < v_R ,
\label{eq:noresto_inf}
\end{equation}
so that the Pati--Salam symmetry is not restored during inflation. After inflation the Universe is described by the Standard Model gauge group, with the right-handed neutrinos present in the spectrum as massive states.

The condition in Eq.~(\ref{eq:noresto_inf}) is not, by itself, sufficient to guarantee that the broken PS phase is maintained throughout the subsequent cosmological evolution. Since the PS-breaking transition occurs before inflation, any topological defects regenerated after inflation would no longer be diluted by subsequent inflation. The relevant requirement is therefore that the post-inflationary plasma never reaches a temperature high enough to restore the PS symmetry.
i.e.
\begin{equation}
v_R \;\gtrsim\; \max\left(T_{\rm RH},\,T_f\right),
\label{eq:noresto_general}
\end{equation}
up to an order-one factor reflecting the ratio of the critical temperature to the VEV in the PS scalar potential. Here $T_{\rm RH}$ is the reheating temperature and $T_f$ is the plasma temperature at PBH formation. Since the PBH-forming modes re-enter the horizon during radiation domination, formation necessarily postdates the completion of reheating and $T_{\rm RH}>T_f$ in general. In what follows we adopt the minimal assumption that reheating completes shortly before the scale $k_f$ re-enters the horizon, so that $T_{\rm RH}\simeq T_f$ and Eq.~(\ref{eq:noresto_general}) is controlled by the formation temperature; for a more prolonged radiation-dominated era between reheating and formation the bound on $v_R$ strengthens by the factor $T_{\rm RH}/T_f$.
In the cosmological history considered here,  the hottest epoch relevant to the present mechanism is the moment of PBH formation, at which the plasma temperature is given by
\begin{equation} \label{eq:Tf_PS}
T_f \simeq 4.36\times10^{15}~{\rm GeV}
\left(\frac{1~{\rm g}}{M_{\rm in}}\right)^{1/2} \,.
\end{equation}
Following PBH formation, the radiation bath cools as $T\propto a^{-1}$ during radiation domination and subsequently as $T\propto a^{-3/8}$ during the PBH-dominated epoch, before approaching the standard radiation-dominated scaling after evaporation.
To strictly avoid the thermal regeneration of the PS symmetry, the breaking scale must remain larger than this  formation temperature:
\begin{equation}
v_R \;\gtrsim\; T_f
\;\simeq\; 4.36\times10^{15}~{\rm GeV}
\left(\frac{1~{\rm g}}{M_{\rm in}}\right)^{1/2} \,.
\label{eq:noresto_Tf}
\end{equation}

Because $T_{\rm ev}<T_f$ throughout the mass range considered, Eq.~(\ref{eq:noresto_Tf}) automatically implies the corresponding non-restoration condition at evaporation. Using $M_{\rm in} = 4\pi\gamma M_p^2/H_f$ with $\gamma\simeq0.2$, the Hubble rate at formation is
\begin{equation}
H_f = \frac{4\pi\gamma M_p^2}{M_{\rm in}}
\simeq 2.7\times10^{13}~{\rm GeV}\left(\frac{1~{\rm g}}{M_{\rm in}}\right) ,
\label{eq:Hf_PS}
\end{equation}
so that $T_f/H_f \simeq 1.6\times10^{2}\,(M_{\rm in}/1\,{\rm g})^{1/2}$. Since $H_{\rm inf}\ge H_f$, the constraint $v_R > H_{\rm inf}$ is weaker than Eq.~(\ref{eq:noresto_Tf}) throughout the mass range $0.4~{\rm g}\lesssim M_{\rm in}\lesssim9.7\times10^{8}$ g permitted by the CMB and BBN bounds \cite{Carr:2020gox}. We note in passing that the lower end of that range is itself the statement $H_f < 2.5\times10^{-5}M_p$, which Eq.~(\ref{eq:Hf_PS}) reproduces as $M_{\rm in}\gtrsim0.4$ g.
Equation~(\ref{eq:noresto_Tf}) is a lower bound on the symmetry-breaking scale that tightens as the PBH mass decreases. For the benchmark $M_{\rm in}\simeq1.2\times10^{6}$ g relevant to remnant dark matter it requires $v_R\gtrsim4\times10^{12}$ GeV, while for $M_{\rm in}\simeq1.3\times10^{2}$ g it requires $v_R\gtrsim3.8\times10^{14}$ GeV. Sub-gram PBHs push $v_R$ to within an order of magnitude of the Planck scale and are therefore disfavoured within a pre-inflationary breaking scenario.

The heavy right-handed neutrinos governing the generation of the lepton asymmetry
can originate from two distinct channels, each constrained differently by the Pati--Salam architecture. First, Hawking emission from the PBHs is efficient only when the emitted species is light compared with the instantaneous black hole temperature \cite{Hawking:1975vcx},
\begin{equation}
T_{\rm BH} = \frac{M_p^2}{M_{\rm BH}}
\simeq 10^{13}~{\rm GeV}\left(\frac{1~{\rm g}}{M_{\rm BH}}\right) ,
\label{eq:TBH_PS}
\end{equation}
where $M_p\simeq2.435\times10^{18}$ GeV is the reduced Planck mass. Combining Eq.~(\ref{eq:TBH_PS}) with Eq.~(\ref{eq:MN_PS}), the natural kinematic threshold for unsuppressed emission,
\begin{equation}
T_{\rm BH}(M_{\rm in}) \gtrsim {M_{N_1}} = y_R v_R ,
\label{eq:emission_PS}
\end{equation}
translates into
\begin{equation} 
M_{\rm in} \lesssim \frac{M_p^2}{y_R v_R}
\simeq 10^{2}~{\rm g}
\left(\frac{10^{11}~{\rm GeV}}{y_R v_R}\right) .
\label{eq:Mmax_PS}
\end{equation}
We emphasise that Eq.~(\ref{eq:Mmax_PS}) is an efficiency criterion and not a hard kinematic cut. Emission above threshold is Boltzmann suppressed rather than forbidden, and this suppression is retained exactly in our numerical treatment through the function $\mathcal{F}(z)$ of Eq.~(\ref{eq:PBHdecay}) with $z=M_{N_1}/T_{\rm BH}$.

Second, if efficient thermal production of $N_1$ is to occur, 
it demands that the plasma reach a temperature comparable to the {RHN} mass {\it i.e.} $M_{N_1}\lesssim T_f$.  Combined with the non-restoration requirement in Eq.~(\ref{eq:noresto_Tf}),  $v_R\gtrsim T_f$, these conditions firmly bracket the Yukawa coupling of the PS scalar sector:
{
\begin{equation}
y_R = \frac{M_{N_1}}{v_R}
\;\lesssim\; \frac{M_{N_1}}{T_f}
\;\simeq\; 2.3\times10^{-5}
\left(\frac{M_{N_1}}{10^{11}~{\rm GeV}}\right)
\left(\frac{M_{\rm in}}{1~{\rm g}}\right)^{1/2} .
\label{eq:yR_bound}
\end{equation}
}
The content of Eq.~(\ref{eq:yR_bound}) lies in its numerical value at the benchmarks rather than in the parametric statement $y_R\lesssim1$, which follows trivially from combining $v_R\gtrsim T_f$ with $M_{N_1}\lesssim T_f$. At the two benchmarks considered in this work it gives $y_R\lesssim0.11$ for $M_{\rm in}\simeq1.2\times10^{6}$ g with $M_{N_1}\simeq4.4\times10^{11}$ GeV, and $y_R\lesssim6.5\times10^{-4}$ for $M_{\rm in}\simeq1.3\times10^{2}$ g with $M_{N_1}\simeq2.5\times10^{11}$ GeV. We note that the requirement $M_{N_1}\lesssim T_f$ applies only when the thermal channel is invoked; in the PBH-only limit it should be dropped, and $y_R$ is then bounded only by the ratio $M_{N_1}/v_R$ with $M_{N_1}$ a free parameter.
The {RHN} mass is thus required to sit well below the $SU(2)_R$ breaking scale, which is a genuine prediction of embedding the mechanism in a pre-inflationary Pati--Salam cosmology rather than an assumption. In presenting numerical results we therefore treat $y_R$ as a derived quantity fixed by the pair $(M_{\rm in}, M_{N_1})$, and not as a free parameter set to unity.

Two further conditions delimit the scenario. The first concerns the initial PBH abundance $\beta=\rho_{\rm BH}/\rho_{\rm rad}|_{T_f}$. Throughout this work we restrict attention to histories in which the black holes come to dominate the total energy density before evaporating, $\beta\ge\beta_{\rm min}$ with $\beta_{\rm min}$ given in Eq.~(\ref{eq:betamin}). We emphasise that this is a choice of cosmological history rather than a consistency requirement of the model. It is, however, the case in which the ambient entropy of the post-evaporation plasma is generated entirely by the black holes themselves rather than inherited from the primordial inflationary bath, so that the reheating scale $T_{\rm ev}|_{\rm MD}$ is fixed purely by the black hole lifetime and the relic abundance of Planck-scale remnants becomes independent of $\beta$. The complementary regime $\beta<\beta_{\rm min}$, in which evaporation occurs during radiation domination, is not excluded and is revisited in Sec.~\ref{sec:dm}, where it is shown to provide one of the two ways of relieving the tension between remnant dark matter and sphaleron reprocessing. Since $\beta_{\rm min}\propto M_{\rm in}^{-1}$, domination becomes progressively easier as the PBH mass is increased, so that the heavier PBHs required by the remnant dark matter condition demand the least fine-tuned curvature enhancement.
The second is that the mass loss driving the evaporation, computed from the Hawking spectrum with greybody corrections \cite{Page:1976df,MacGibbon:1991tj}, terminate early enough for the accumulated $(B-L)$ asymmetry to be reprocessed by the electroweak sphalerons. When the thermal channel contributes, the asymmetry is already present in the plasma well before evaporation and this restriction is correspondingly relaxed.

Table~\ref{tab:benchmarks} collects the two benchmark points used throughout this work and the values taken by each of the conditions discussed above. BP1 realizes the PBH-only leptogenesis limit; BP2 is the mass required for Planck-scale remnants to saturate the observed dark matter abundance, and fails the sphaleron condition, as anticipated in Sec.~\ref{sec:intro} and quantified in Sec.~\ref{sec:dm}.

\begin{table}[t]
\centering
\begin{tabular}{lcc}
\hline\hline
 & BP1 & BP2 \\
\hline
$M_{\rm in}$ [g] & $1.3\times10^{2}$ & $1.2\times10^{6}$ \\
$M_{N_1}$ [GeV] & $2.5\times10^{11}$ & $4.4\times10^{11}$ \\
$\beta$ & $4.2\times10^{-6}$ & $2.7\times10^{-10}$ \\
\hline
$T_f$ [GeV], Eq.~(\ref{eq:Tf_PS}) & $3.8\times10^{14}$ & $4.0\times10^{12}$ \\
$v_R \gtrsim$ [GeV], Eq.~(\ref{eq:noresto_Tf}) & $3.8\times10^{14}$ & $4.0\times10^{12}$ \\
$y_R \lesssim$, Eq.~(\ref{eq:yR_bound}) & $6.5\times10^{-4}$ & $0.11$ \\
$\beta_{\rm min}$, Eq.~(\ref{eq:betamin}) & $5.2\times10^{-8}$ & $5.7\times10^{-12}$ \\
\hline
PBH domination & satisfied & satisfied \\
Sphaleron bound, Eq.~(\ref{eq:Mmax_PS}) & satisfied & violated \\
\hline\hline
\end{tabular}
\caption{Cosmological consistency conditions of Sec.~\ref{sec:PS} evaluated at the two benchmark
points used in the numerical analysis.}
\label{tab:benchmarks}
\end{table}

Collecting the above, the Pati--Salam embedding correlates the symmetry-breaking scale with the PBH mass through the interdependent relations:
\begin{equation}
4.36\times10^{15}~{\rm GeV}\left(\frac{1~{\rm g}}{M_{\rm in}}\right)^{1/2}
\;\lesssim\; v_R ,
\qquad
M_{N_1} = y_R v_R \;\lesssim\; T_f ,
\qquad
0.4~{\rm g}\;\lesssim\; M_{\rm in},
\end{equation}
so that the breaking scale, the inflationary scale, and the PBH mass are not independent inputs. The structure of the allowed region is controlled by the single relation $v_R \gtrsim T_f \propto M_{\rm in}^{-1/2}$, which permits low breaking scales only for heavy PBHs, precisely the regime selected by the remnant dark matter requirement analyzed later in this work. Within this window $v_R$ is constrained by cosmological consistency and by the leptogenesis requirements rather than by specific gauge unification assumptions \cite{Mohapatra:1986uf,Babu:1992ia,Bajc:2006ia,Bertolini:2009es}. The Pati--Salam structure therefore determines the kinematically consistent region and ties the RHN mass to a gauge symmetry breaking scale, but does not by itself fix the normalisation of the baryon asymmetry.

\section{Inflation and primordial black hole formation}
\label{sec:inflation}

We assume that PBHs originate from enhanced curvature perturbations generated during inflation. The inflaton sector consists of a single real gauge singlet scalar field $\phi$, minimally coupled to gravity, with canonical kinetic term and potential $V(\phi)$. Being a gauge singlet, $\phi$ is inert under $G_{\rm PS}$ and therefore plays no role in the symmetry-breaking chain of Eq.~(\ref{eq:chain}) and the Pati-Salam transition is assumed to have completed before the onset of the inflationary phase described here, as discussed in Sec.~\ref{sec:PS}. To make the origin of the required small-scale enhancement explicit, we consider a potential possessing a localized near-inflection feature, a construction widely employed to generate PBHs within single-field inflation \cite{Garcia-Bellido:2017mdw,Germani:2017bcs,Motohashi:2017kbs,Ballesteros:2017fsr,Hertzberg:2017dkh}. Expanding about the feature at $\phi=\phi_0$,
\begin{equation}
V(\phi)
=
V_0
\left[
1
+
a\,\frac{\phi-\phi_0}{M_p}
+
b\left(\frac{\phi-\phi_0}{M_p}\right)^{3}
\right],
\label{eq:inflaton_potential}
\end{equation}
where $V_0$ sets the overall inflationary energy density, $\phi_0$ denotes the approximate inflection point, and $a$ and $b$ are dimensionless parameters controlling deviations from exact flatness. Equation~(\ref{eq:inflaton_potential}) is a local expansion valid in the neighbourhood of the feature. The global completion of the potential, which must possess a minimum in which inflation terminates and reheating proceeds, is model dependent and plays no role in the small-scale enhancement studied here; explicit realizations may be found in Refs.~\cite{Garcia-Bellido:2017mdw,Ballesteros:2017fsr,Hertzberg:2017dkh}. Since Eq.~(\ref{eq:inflaton_potential}) contains no quadratic term, $V''(\phi_0)=0$ holds identically, and the near-inflection condition
\begin{equation}
V'(\phi_0) \simeq 0,
\qquad
V''(\phi_0) \simeq 0,
\end{equation}
reduces to the single requirement $|a|\ll1$, leading to a transient ultra-slow-roll phase during which the inflaton velocity $\dot{\phi}$ is strongly suppressed.
Since $V'(\phi_0)=V_0\,a/M_p$ and $V(\phi_0)=V_0$, the potential slow-roll parameter,, which controls the inflaton kinetic energy through $\dot{\phi}^2 \simeq 2\epsilon_V V$, at the feature is fixed directly by $a$,
\begin{equation}
\epsilon_V(\phi_0) = \frac{M_p^2}{2}\left(\frac{V'}{V}\right)^2_{\phi_0} = \frac{a^2}{2},
\label{eq:epsV_a}
\end{equation}
so that the parameter $a$ is not merely small but quantitatively determined by the required amplitude of the enhancement, as we make explicit below. The cubic coefficient $b$ controls the width of the feature in field space and hence the width of the resulting peak in $k$-space, and is chosen such that the enhancement is narrow enough for the monochromatic approximation adopted here.

The gauge-invariant curvature perturbation $\zeta$, defined on uniform-density hypersurfaces, has dimensionless power spectrum
\begin{equation}
\mathcal{P}_\zeta(k)
=
\frac{k^3}{2\pi^2}
\langle |\zeta_k|^2 \rangle,
\end{equation}
where $k$ is the comoving wavenumber and $\zeta_k$ denotes the Fourier mode. In standard slow-roll inflation this spectrum is evaluated at horizon exit $k=aH$, approximately~\cite{Mukhanov:1990me,Lidsey:1995np,Lyth:1998xn,Baumann:2009ds}
\begin{equation}
\mathcal{P}_\zeta(k)
\simeq
\frac{1}{24\pi^2 M_{p}^4}
\frac{V}{\epsilon_V}\,.
\label{eq:power_spectrum}
\end{equation}
%
We use Eq.~(\ref{eq:power_spectrum}) only as an order-of-magnitude estimate of the enhancement. In the vicinity of an inflection point the slow-roll hierarchy is by construction violated, and an accurate spectrum requires numerical solution of the Mukhanov--Sasaki equation \cite{Motohashi:2017kbs,Hertzberg:2017dkh}, which typically shifts the peak amplitude by an $\mathcal{O}(1)$ to $\mathcal{O}(10)$ factor. Because the PBH abundance depends on $\mathcal{P}_\zeta$ only logarithmically, this does not affect the conclusions drawn below.

Near the inflection point the suppression of $V'$ reduces $\epsilon_V$ by several orders of magnitude while $V\simeq V_0$ remains essentially constant across the feature. A temporary reduction from $\epsilon_V \sim 10^{-3}$, consistent with the bound $r \simeq 16\,\epsilon_V < 0.036$ on CMB scales \cite{Planck:2018jri, BICEP:2021xfz}, to $\epsilon_V \sim 10^{-10}$ enhances the curvature spectrum from
\begin{equation}
\mathcal{P}_\zeta(k_*) \simeq 2.1\times10^{-9}
\end{equation}
at the CMB pivot scale $k_* = 0.05\,{\rm Mpc}^{-1}$ \cite{Planck:2018jri} to
\begin{equation}
\mathcal{P}_\zeta(k_f) \sim 10^{-2}
\end{equation}
at the small scale $k_f$ associated with PBH formation. Because the feature is localized in field space, this enhancement is confined to ultraviolet scales and leaves large-scale observables unaffected.

After inflation, enhanced modes re-enter the Hubble horizon during radiation domination. We assume throughout that reheating following inflation completes before the scale $k_f$ re-enters the horizon, so that formation occurs in a radiation-dominated background; as noted in Sec.~\ref{sec:PS}, we take the minimal case in which the two epochs are close, $T_{\rm RH}\simeq T_f$, since a prolonged radiation era between reheating and formation would strengthen the non-restoration bound of Eq.~(\ref{eq:noresto_general}) by the factor $T_{\rm RH}/T_f$. The comoving curvature perturbation $\zeta$ is related to the density contrast $\delta$ at horizon crossing by $\delta \simeq (4/9)\zeta$ in radiation domination. Regions collapse to form PBHs if their density contrast exceeds a threshold $\delta_c \sim 0.4$ \cite{Carr:1975qj,Musco:2004ak,Harada:2013epa}. 

The initial PBH mass at the formation temperature $T_f$ is approximately related to the horizon mass at re-entry,
\begin{equation}\label{eq:MassPBH}
M_{\rm in}
\simeq
\gamma
\frac{4\pi}{3}
\frac{\rho_{\rm tot}(T_f)}{H^3(T_f)},
\end{equation}
where the efficiency of collapse is $\gamma\simeq 0.2$~\cite{Carr:1975qj,Green:2004wb} and the total energy density at the formation temperature is $\rho_{\rm tot}(T_f) = 3 M_p^2 {H}^2(T_f)$, so that Eq.~(\ref{eq:MassPBH}) reduces to $M_{\rm in}=4\pi\gamma M_p^2/H_f$ as used in Eq.~(\ref{eq:Hf_PS}). Therefore, the corresponding formation temperature $T_f$ is related to $M_{\rm in}$ as: 
\begin{equation}\label{eq:initialTem}
    T_f =\left(\frac{1440 \ \gamma^2}{g_*(T_f)}\right)^{1/4} M_p \sqrt{\frac{M_p}{M_{\rm in}}}\simeq 4.36 \times 10^{15}~\text{GeV} \left(\frac{1~\rm{g}}{M_{\rm in}}\right)^{1/2}.
\end{equation}
For this calculation, the PBH formation time $t_{f}$ is determined by 
\begin{equation} \label{eq:PBHformation_time} 
t_{f} = \frac{M_{\rm in}}{8\pi \gamma M^2_p}. 
\end{equation}
Equivalently, the comoving scale associated with mass $M_{\rm in}$ becomes
\begin{equation}\label{eq:k_f}
    k_f = a_f H_f \simeq 7.75 \times 10^{22} \ {\rm{Mpc}^{-1}} \left(\frac{1~\rm{g}}{M_{\rm in}}\right)^{1/2} \simeq 2.45 \times 10^{15} \ {\rm{Mpc}^{-1}} \left(\frac{10^{15}~\rm{g}}{M_{\rm in}}\right)^{1/2}\,,
\end{equation}
with the corresponding scale factor given by 
\dis{
\frac{a_f}{a_0} \simeq \frac{T_0}{T_f} \left(\frac{g_{*,s}(T_0)}{g_{*,s} (T_f)} \right)^{1/3} \simeq 1.8 \times 10^{-29} {\left(\frac{M_{\rm in}}{1~\rm{g}}\right)^{1/2}}\,.
}
From Eq.~(\ref{eq:k_f}), one can realize that the comoving scale $k_f$
lies many orders of magnitude {beyond the range of scales probed by the CMB.}

The initial mass fraction at formation is
\begin{equation}
\beta({M_{\rm in}})
=
\int_{\delta_c}^{\infty}
\frac{1}{\sqrt{2\pi}\sigma({M_{\rm in}})}
\exp\!\left(-\frac{\delta^2}{2\sigma^2({M_{\rm in}})}\right)
d\delta,
\end{equation}
where $\sigma^2({M_{\rm in}})$ is the variance of the smoothed density contrast. For Gaussian perturbations and $\sigma \ll \delta_c$, the integral gives
\begin{equation}
\beta({M_{\rm in}})
\approx
\frac{\sigma({M_{\rm in}})}{\delta_c \sqrt{2\pi}}
\exp\!\left(-\frac{\delta_c^2}{2\sigma^2({M_{\rm in}})}\right).
\label{eq:beta_asymptotic}
\end{equation}
The variance is related to the curvature spectrum through
\begin{equation}
\sigma^2({M_{\rm in}})
\sim
\frac{16}{81}
\mathcal{P}_\zeta(k_f),
\label{eq:sigma_Pzeta}
\end{equation}
up to order-one window-function factors, the numerical coefficient following from $\delta\simeq(4/9)\zeta$ for a narrow peak. Equations~(\ref{eq:beta_asymptotic}) and (\ref{eq:sigma_Pzeta}) determine the curvature amplitude required to realize a given initial abundance. The relevant range of $\beta$ in this work is set at the lower end by the PBH-domination requirement of Eq.~(\ref{eq:betamin}), and at the upper end by the requirement that the induced gravitational wave background not spoil the effective number of relativistic species at nucleosynthesis, discussed in Sec.~\ref{sec:gw}.

Achieving $\beta \sim 10^{-8}$--$10^{-7}$ therefore requires $\sigma \sim 0.05$--$0.1$, corresponding to
\begin{equation}
\mathcal{P}_\zeta(k_f) \sim \mathcal{O}(10^{-2}).
\end{equation}

Although this amplitude is several orders of magnitude larger than the CMB value, it is phenomenologically viable provided the enhancement is sufficiently narrow in $k$-space. Narrow features avoid constraints from CMB spectral distortions and pulsar timing arrays \cite{Chluba:2012we,Inomata:2019zqy}, and are well approximated by a monochromatic PBH mass function. In the present work we adopt this monochromatic approximation, which captures the dominant dynamics and allows a transparent analytic connection between inflationary parameters, PBH mass, {evaporation} temperature, and the baryon asymmetry derived in subsequent sections.


\section{PBH evaporation and non-resonant leptogenesis}
\label{sec:leptogenesis}
After formation, PBHs can emit all particles in the spectrum through Hawking evaporation provided their masses satisfy $m_i \lesssim T_{\rm BH}$ where 
\begin{equation}\label{eq:Hawkingtemperature}
    T_\text{BH} = \frac{M_p^2}{M_\text{BH}} \simeq 10^{13}~\rm GeV \left(\frac{1~\rm g}{M_{\rm BH}}\right)\,,
\end{equation}
with the reduced Planck mass is $M_p \simeq 2.435 \times 10^{18}~\rm{GeV} \simeq 4.33 \times 10^{-6} \ \rm{g}$. 
For a Schwarzschild black hole, the greybody factor in the geometric optics (GO) limit is approximated as $\sigma_{s_i} = (27/64\pi) M_{\rm BH}^2/M_p^4$. Consequently, the spectrum simplifies to~\cite{UKWATTA201690,Lunardini:2019zob,Perez-Gonzalez:2020vnz}  
\begin{equation}
    \frac{d^2 u_{i}}{dtdE} \simeq \frac{27 g_i}{128  \pi^3} \frac{M_{\rm BH}^2}{M_p^4} \frac{E_i^3}{e^{E_i/T_{\rm BH}}-(-1)^{2s_i}}\,.
\end{equation}  
with the sign in the denominator distinguishing bosonic from fermionic emission.
Defining the comoving energy densities of radiation and PBH as $\tilde{\rho}_r = a^4 \rho_r$ and $\tilde{\rho}_{\rm BH} = a^3 \rho_{\rm BH}$, the Boltzmann equations governing their evolution are~\cite{Masina:2020xhk,Giudice:2000ex,Bernal:2020bjf,JyotiDas:2021shi,Barman:2021ost,Choi:2023kxo}  
\begin{equation}
\begin{aligned}\label{eq:Bolteq}
    & \frac{dM_\text{BH}}{d \ln(a)} = - \frac{\mathcal{E}(M_\text{BH})}{{H}} \frac{ M_p^4}{M_\text{BH}^2}, \\
    & \frac{d \tilde{\rho}_\text{BH}}{d \ln(a)} =  \frac{\tilde{\rho}_\text{BH}}{M_\text{BH}} \frac{dM_\text{BH}}{d \ln(a)}, \\
    & \frac{d \tilde{\rho}_r}{d \ln(a)} = -\frac{\mathcal{E}_\text{SM}(M_\text{BH})}{\mathcal{E}(M_\text{BH})} \frac{a \ \tilde{\rho}_\text{BH}}{M_\text{BH}} \frac{dM_\text{BH}}{d \ln(a)},
\end{aligned}
\end{equation}
where {$\mathcal{E}_{\rm SM}\equiv \sum_{i\in {\rm SM}} g_{i}\mathcal{E}_i (z_i)$} accounts for the evaporation contributions from Standard Model particles, excluding right-handed neutrinos. {The total PBH evaporation function is $\mathcal{E}(M_\text{BH}) \equiv \sum_i g_i \mathcal{E}_i (z_i)$ with $z_i=m_i/T_{\rm BH}$ and}
\begin{equation}\label{eq:evap Func}
    \mathcal{E}_i (z_i) =\frac{27}{128\pi^3} \int_{z_i}^\infty \frac{\psi_{s_i}(x) (x^2 - z_i^2)}{e^x - (-1)^{2s_i}} x dx.
\end{equation}
{We denote the evaporation function by $\mathcal{E}$ rather than $\varepsilon$ to avoid confusion with the CP asymmetry parameter $\varepsilon_1$ introduced in Sec.~\ref{sec:lepto_sub}.}
In the GO limit ($\psi_{s_i}=1$), PBH mass evolution follows
\begin{equation}\label{eq:PBHmass}
    M_\text{BH}(t) = M_\text{in} \left(1- \frac{t-t_{i}}{\tau_{\rm BH}}\right)^{1/3},
\end{equation}
with the PBH lifetime given by
\begin{equation}\label{eq:lifetime}
    \tau_{\rm BH}= \frac{4}{27} \frac{160 \ M_\text{in}^3}{\pi \ g_*(T_\text{BH}) M_p^4}  \simeq 2.66\times 10^{-28}~\rm{s}~\frac{100}{g_*(T_\text{BH})}\left(\frac{M_{\rm in}}{1~\rm{g}}\right)^{3}.
\end{equation}
Since most of the PBH energy is emitted near the end of evaporation, the evaporation temperature, $T_\text{ev}$, depends on whether the Universe is PBH dominated (matter-dominated, MD) or radiation-dominated (RD) prior to evaporation
\begin{equation}\label{eq:evapTemperature}
  T_\text{ev} |_\text{MD} \simeq 3.55 \times 10^{10}~\text{GeV} \left(\frac{1~\text{g}}{M_\text{in}}\right)^{3/2},
\end{equation}
\begin{equation}\label{eq:ev temperaturePBH}
    T_\text{ev} |_\text{RD} \simeq \frac{\sqrt{3}}{2} T_\text{ev} |_\text{MD}\,,
\end{equation}
{the ratio following from $H_{\rm ev}=1/2\tau_{\rm BH}$ during radiation domination as against $H_{\rm ev}=2/3\tau_{\rm BH}$ during matter domination.}
An important quantity is the ratio of the PBH lifetime to the formation time, $t_f$, defined in Eq.~(\ref{eq:PBHformation_time}),
\dis{
\frac{\tau_{\rm BH}}{t_{f}} \simeq \frac{38}{g_*(T_{\rm BH})} \left(\frac{M_{\rm in}}{M_p}\right)^2 \gg 1\,,
}
indicating that the PBH lifetime is much longer than the formation time. Consequently, an early PBH dominated era can be readily realized. The evaporation time is given by $t_{\rm ev}= \tau_{\rm BH} + t_f$, which can be approximated as $t_{\rm ev} \simeq \tau_{\rm BH}$, since $t_{f} \ll \tau_{\rm BH}$. If the Universe indeed undergoes an early MD epoch driven by PBHs, we denote by $a_{\rm eeq}$ the scale factor at the epoch of early radiation--PBH equality. At the early equality time and the PBH formation time, the energy densities satisfy
\dis{
3 M_p^2 H^2_{\rm eeq} = \rho_{\rm BH}(T_{\rm eeq}) + \rho_r(T_{\rm eeq}) = 2 \rho_{\rm BH}(T_{\rm eeq})\,,
}
and 
\dis{
3 M_p^2 H^2_f = \rho_{\rm BH}(T_f) + \rho_r (T_f) =\rho_{\rm BH}(T_{\rm eeq}) \left[ \left(\frac{a_f}{a_{\rm eeq}}\right)^{-3} + \left(\frac{a_f}{a_{\rm eeq}}\right)^{-4}\right]\,.
}
Comparing them, one can obtain 
\dis{\label{eq:ratio_early_equality}
\frac{H^2_{\rm eeq}}{H^2_f} = \frac{2}{\left[ \left(\frac{a_f}{a_{\rm eeq}}\right)^{-3} + \left(\frac{a_f}{a_{\rm eeq}}\right)^{-4}\right]} \approx 2 \left(\frac{a_f}{a_{\rm eeq}}\right)^{4}\,.
}
Thus, the definition of $\beta$ can be rewritten as 
\dis{
\beta = \frac{\rho_{\rm BH} (T_f)}{\rho_{\rm tot}(T_f)} = \frac{\rho_{\rm BH}(T_{\rm eeq}) \left(\frac{a_f}{a_{\rm eeq}}\right)^{-3}}{\rho_{\rm BH}(T_{\rm eeq}) \left[ \left(\frac{a_f}{a_{\rm eeq}}\right)^{-3} + \left(\frac{a_f}{a_{\rm eeq}}\right)^{-4}\right]} = \frac{1}{1+\left(\frac{a_f}{a_{\rm eeq}}\right)^{-1}}\,,
}
which is simplified to $1/\beta = 1+ \left(\frac{a_{\rm eeq}}{a_f}\right)$. If $\beta\ll 1$, Eq.~(\ref{eq:ratio_early_equality}) becomes 
\dis{\label{eq:HubbleRatio_wrt_beta}
\frac{H_{\rm eeq}}{H_f} \approx \sqrt{2} \beta^2\,.
}
Therefore, depending on the initial PBH energy density fraction, PBH evaporation may occur either during the RD epoch or during the early PBH-driven MD epoch. Imposing the condition for the existence of the PBH dominated era, namely $H_{\rm ev}\leq H_{\rm eeq}$, the minimum initial PBH abundance is obtained as
\dis{\label{eq:betamin}
\beta_{\rm min} \simeq \left(\frac{H_{\rm ev}}{ \sqrt{2} H_f}\right)^{1/2} &= \frac{1}{2^{1/4}} \frac{\left.T_{\rm ev}\right|_{\rm MD}}{T_f} = 7 \times 10^{-2}  \left(\frac{g_*(T_{\rm BH})}{\gamma}\right)^{1/2} \frac{M_p}{M_{\rm in}} \\
    & \simeq 6.8 \times 10^{-6} \left(\frac{0.2}{\gamma}\right)^{1/2} \left(\frac{g_*(T_{\rm BH})}{100}\right)^{1/2}\left(\frac{1 \ \rm g}{M_{\rm in}}\right).
} 
Hence, a $1 \ \rm{g}$ PBH dominates the energy density before evaporation if $\beta \gtrsim 6.8 \times 10^{-6}$.

As we assume a monochromatic PBH mass spectrum, the comoving PBH number density remains conserved:  
\begin{equation}\label{eq:nbhconst}
    n_{\rm BH} (t) = n_{\text{BH}}(t_f) \left(\frac{a_f}{a}\right)^3,
\end{equation}
where the initial number density $n_{\rm BH}(t_f)$ is given by  
\begin{equation}\label{eq:nBHin}
    n_{{\rm BH}} (t_f) = \beta  \frac{\rho_{\rm r} (t_f)}{M_{\rm in}} = \beta \frac{48 \pi^2 \gamma^2 M_p^6}{M_{\rm in}^3}.
\end{equation}
 In addition to Eq.~(\ref{eq:Bolteq}), the temperature evolution can be written in a simple form by taking into account the non-conservation of entropy due to PBH evaporation
\dis{\label{eq:Tem_Boltz}
\frac{dT}{d \ln (a)} = - {T} \left(1+ \frac{T}{3 g_{*,s}(T)} \frac{d g_{*,s} (T)}{{dT}} \right)^{-1} \left[1- \frac{1}{4} \frac{g_*(T)}{g_{*,s}(T)} \frac{1}{\tilde{\rho}_{\rm rad}} \frac{d \tilde{\rho}_{\rm rad}}{d \ln(a)} \right]\,.
}
{Equation~(\ref{eq:Tem_Boltz}) is the relation that carries the entropy injected by the black holes into the plasma, and it is through this equation that the dilution of any pre-existing asymmetry is captured; we return to this point in Sec.~\ref{sec:lepto_sub}. Throughout the numerical analysis we adopt $g_*(T_{\rm BH})=g_{*,s}(T_{\rm BH})=106.75$ and $g_{N_1}=2$.}

\begin{figure}[ht]
\centering
\includegraphics[width=0.49\textwidth]{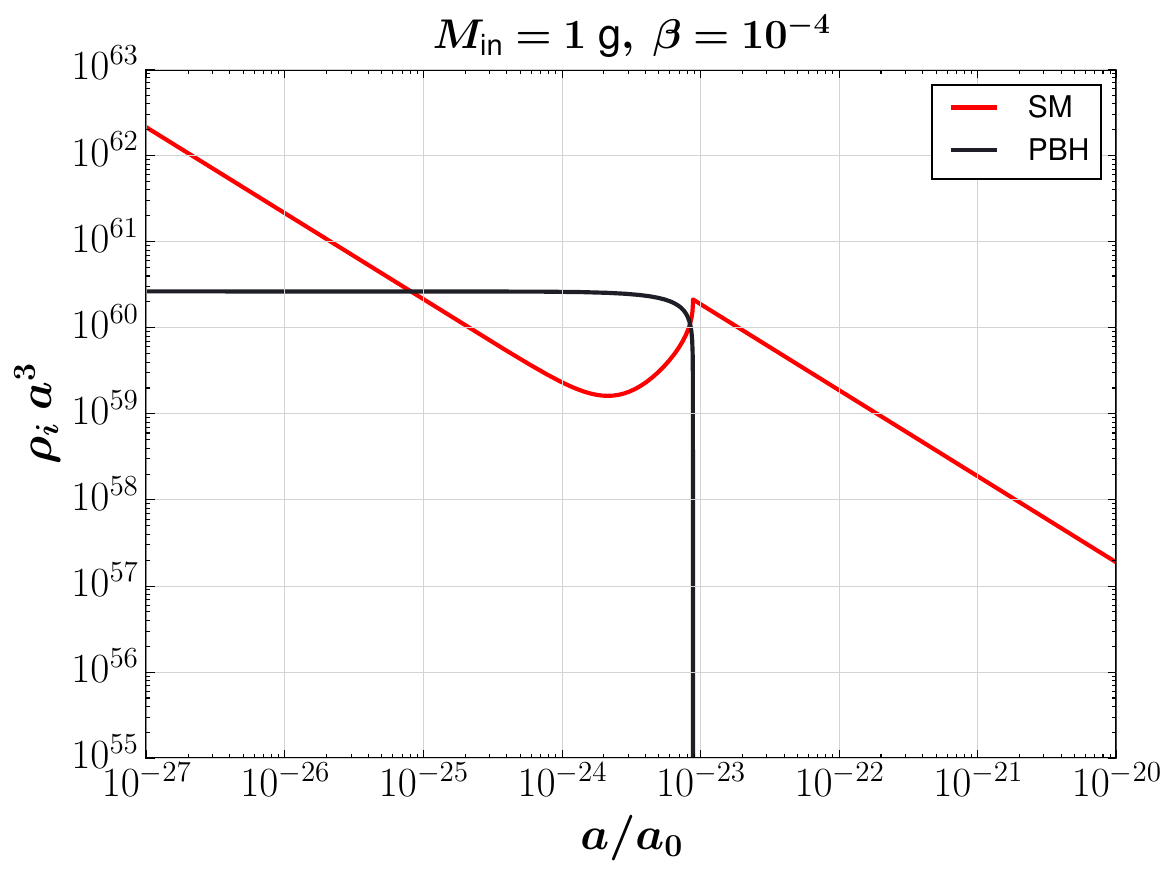}
\hfill
\includegraphics[width=0.49\textwidth]{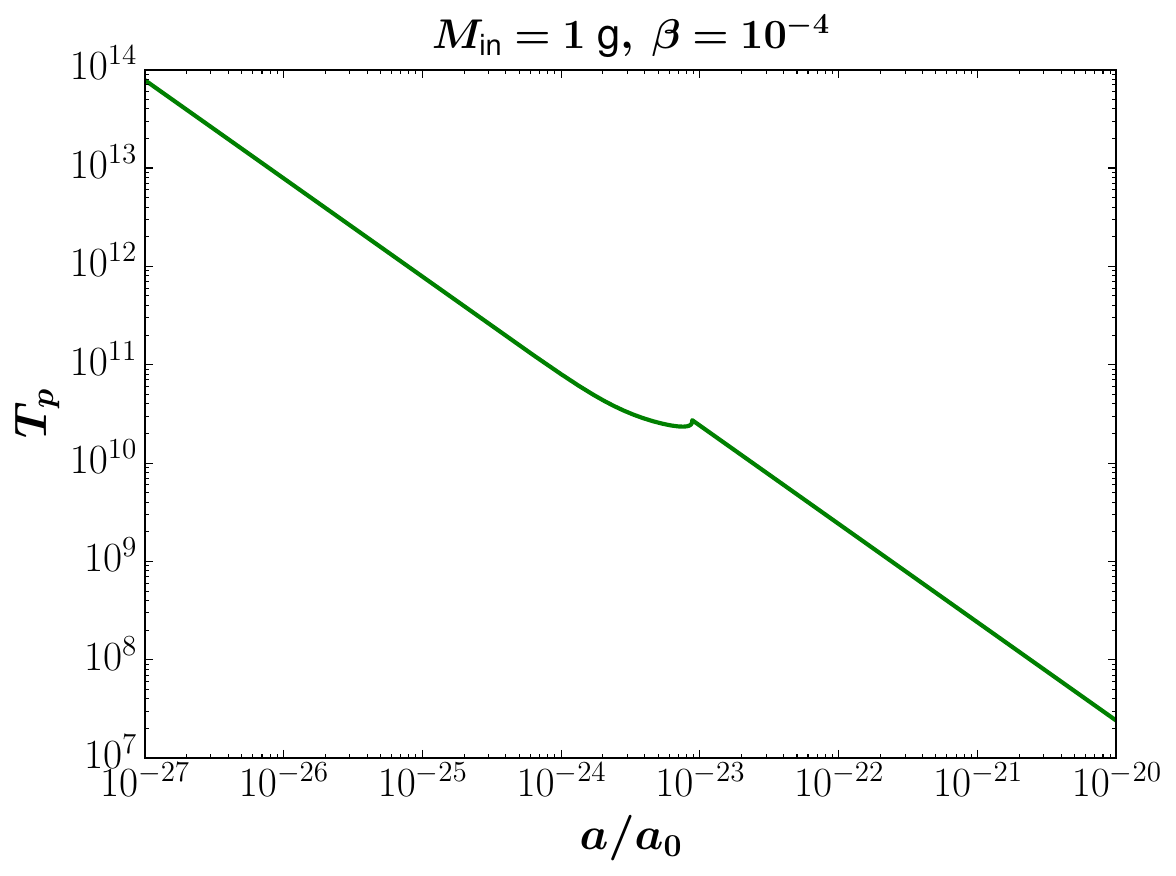}
\hfill
\includegraphics[width=0.49\textwidth]{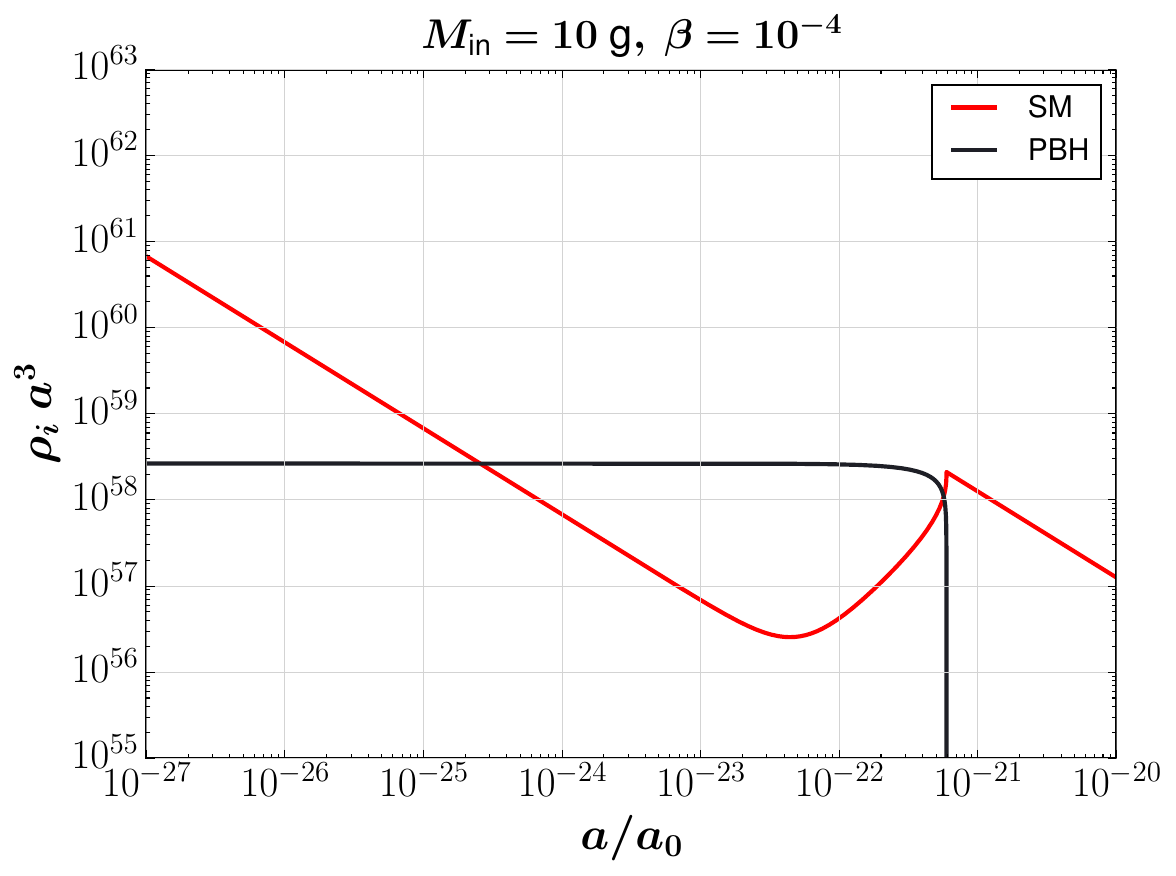}
\hfill
\includegraphics[width=0.49\textwidth]{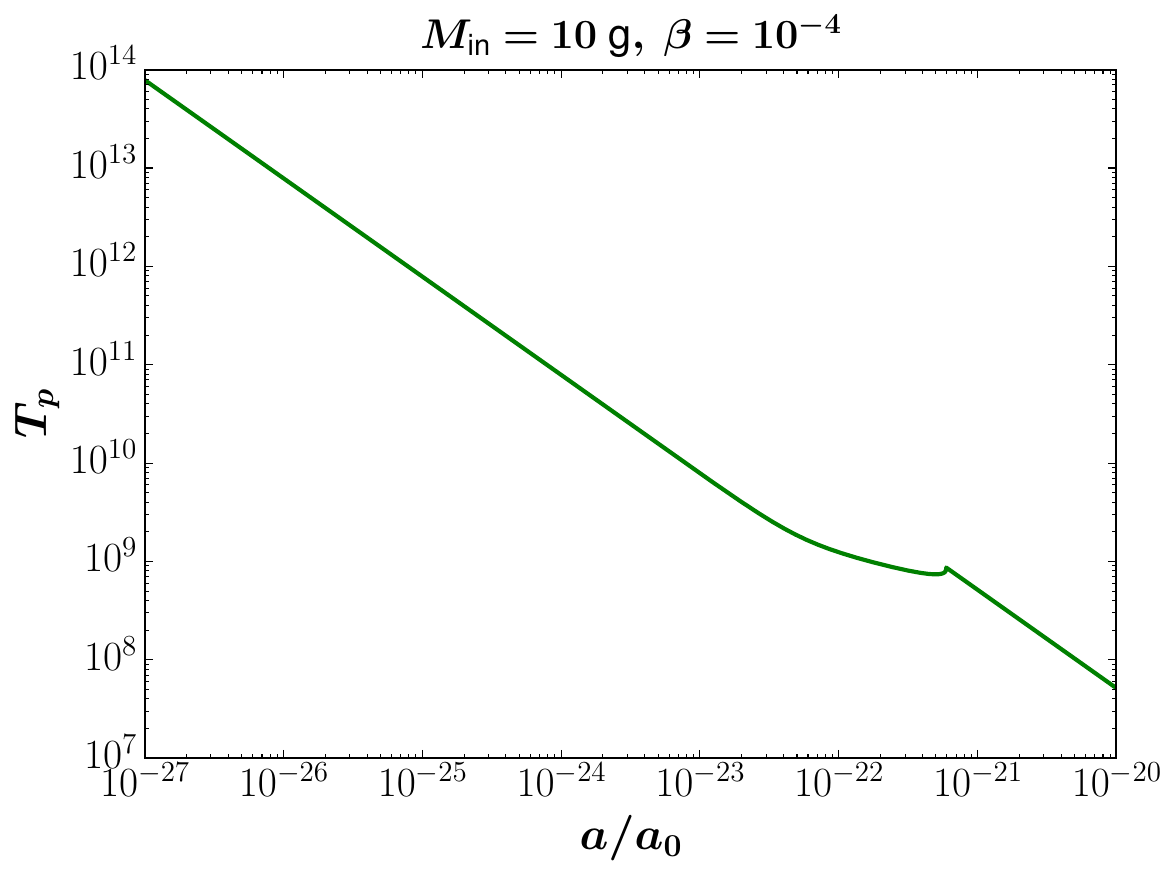}
\caption{
Evolution of the comoving energy densities for radiation and PBH (left), together with the plasma temperature (right) as functions of the scale factor for two different initial mass of PBHs, $M_{\rm in} = 1\ \rm{g}$ and $10 \ \rm{g}$. In both cases, the initial PBH abundance is fixed to $\beta=10^{-4}$, which leads to an early PBH-induced MD epoch. 
}
\label{fig:rho_evolution}
\end{figure}

In Fig.~\ref{fig:rho_evolution}, we show the numerical results obtained by simultaneously solving the Boltzmann equations in Eq.~(\ref{eq:Bolteq}) and the plasma temperature evolution equation, Eq.~(\ref{eq:Tem_Boltz}). The left panels show the evolution of the comoving energy densities of radiation (red), and PBH (black) as functions of the scale factor $a$. For the chosen initial conditions, PBHs come to dominate the energy density of the Universe before completely evaporating. Their energy density shows a sharp decrease at $a/a_0\sim 10^{-23}$ in the top left panel for the initial mass of PBHs $M_{\rm in}=1\ \rm{g}$ with $\beta=10^{-4}$, and at $a/a_0\sim 10^{-21}$ in the bottom left panel for $M_{\rm in}=10 \ \rm{g}$ with the same value of $\beta$. 
 Since the PBH energy density redshifts as non-relativistic matter, their comoving energy density remains constant throughout the evolution. During the MD epoch, the radiation energy density scales as $a^{-3/2}$, implying that the comoving radiation energy density grows as $a^{3/2}$. The right panels display the evolution of the plasma temperature as a function of the scale factor $a$. A distinct kink appears in the temperature evolution at approximately $a/a_0\sim 10^{-23}$ (top) and $10^{-21}$ (bottom), denoting the onset of the entropy injection from PBH evaporation into the thermal plasma. Following the completion of PBH evaporation, the Universe returns to the standard RD evolution, and the plasma temperature resumes its usual scaling, $T \propto a^{-1}$.


\subsection{Heavy RHN Production from PBH}
We now compute the baryon asymmetry generated by heavy right-handed neutrinos ($N_1$) emitted during PBH evaporation. The Boltzmann equation tracking the evolution of RHNs produced from PBH evaporation is:  
\begin{equation}\label{eq:boltzeqN} 
     \frac{d \tilde{n}_{N_1}^{\rm BH}}{d \ln(a)} =  \frac{\tilde{\rho}_\text{BH}}{M_\text{BH}} \frac{\Gamma_{\text{BH}\rightarrow N_1}}{{H}} - \frac{{\Gamma_{N_1}^{\rm BH}}}{{H}}  \tilde{n}_{N_1}^{\rm BH},
\end{equation}
which has to be solved along with Eq.~(\ref{eq:Bolteq}) simultaneously.
Here $\tilde{n}_{N_1} \equiv n_{N_1} a^3$ is the comoving number density of RHNs, and $\Gamma_{N_1}^{\rm BH}$ is the decay width corrected by an average inverse time dilatation factor as given by~\cite{Perez-Gonzalez:2020vnz}
\dis{
\Gamma_{N_1}^{\rm BH} \equiv \left\langle \frac{M_{N_1}}{E_{N_1}}\right\rangle_{\rm BH} \Gamma_{N_1} \simeq \frac{K_1(M_{N_1}/T_{\rm BH})}{K_2(M_{N_1}/T_{\rm BH})} \Gamma_{N_1} \,,
}
where the total RHN decay width is $\Gamma_{N_1} ={(y_\nu^\dagger y_\nu) M_{N_1}}/{8\pi}$ with $y_\nu^2
\sim
\frac{m_\nu M_{N_1}}{v^2}$, $m_\nu \sim 0.05$ eV, and $v=174$ GeV.
The momentum-integrated sterile neutrino emission rate from PBH evaporation is:  
\begin{equation}
        \Gamma_{\text{BH}\rightarrow N_1}(t) = \frac{27 g_{N_1}}{128 \pi^3} \frac{M_p^2}{M_\text{BH}(t)}\int_{z}^\infty \frac{\psi_{N_1}(x) (x^2 - z^2)}{e^x - 1}dx,
\end{equation}
where $x=E/T_\text{BH}$ and $z = m_{N_1}/T_{\rm BH}$. In the GO limit, the emission rate simplifies to~\cite{Cheek2022,Perez-Gonzalez:2020vnz}:  
\begin{equation} \label{eq:PBHdecay}
\Gamma_{\text{BH}\rightarrow N_1}(t) = \frac{27 g_{N_1}}{64 \pi^3} \frac{M_p^2}{M_\text{in}} \left(1-\frac{t-t_i}{\tau}\right)^{-1/3}\mathcal{F}(z),
\end{equation}
where $\mathcal{F}(z) = [z  {\rm Li}_2(e^{-z}) + {\rm Li}_3(e^{-z})]$, with ${\rm Li}_n$ denoting the polylog function of order $n$. In the high-temperature limit ($z\ll1$), $\mathcal{F}(z) \to \zeta(3) \approx 1.20206$. For our numerical analysis, we utilize the "\texttt{ULYSSES}" package~\cite{Granelli:2020pim}, incorporating modified graybody factors.

Using the sudden decay approximation with $\rho_{\rm BH}(t_{\rm ev}) = \rho_{\rm rad}(t_{\rm ev})$, we find the relation,
\dis{
\rho_{\rm BH}(t_{\rm ev})= \rho_{\rm BH}(t_f) \left(\frac{a_f}{a_{\rm ev}}\right)^3= \rho_{r}(t_{\rm ev}) = \frac{\pi^2}{30}g_*(T_{\rm ev}) T_{\rm ev}^4\,.
}
The ratio of scale factors $a_f/a_{\rm ev}$ is determined using entropy conservation, along with Eqs.~(\ref{eq:initialTem}) and (\ref{eq:evapTemperature}), and is given by:
\dis{
\frac{a_f}{a_{\rm ev}} = \frac{T_{\rm ev}|_{\rm MD}}{T_{f}} \left(\frac{g_{*,s}(T_{\rm ev})}{g_{*,s}(T_{f})}\right)^{1/3} \simeq 1.9 \frac{M_p}{M_{\rm in}}.
}
Therefore,
\dis{\label{eq:nbhiai3}
n_{\text{BH}}(t_f) \left(\frac{a_f}{a_{\rm ev}}\right)^3 = \frac{\rho_{r}(t_{\rm ev})}{M_{\rm in}}=\frac{ \frac{\pi^2}{30}g_*(T_{\rm ev}) T_{\rm ev}^4}{M_{\rm in}} \simeq 2.33 \times 10^2 \frac{M_p^{10}}{M_{\rm in}^7}.
}
The sterile neutrino number density is estimated as 
\dis{\label{eq:nRHN_GO}
n_{N_1}(t_{\rm ev}) \simeq \frac{15~\zeta(3)g_{N_1} }{\pi^4g_*(T_{\rm BH}) } \frac{M^2_{\rm in}}{M^2_{p}} n_{\rm BH}(t_f) \left(\frac{a_f}{a_{\rm ev}}\right)^3 \simeq 1.9 \times 10^{23} \ \rm{GeV}^3 \left(\frac{10 \ \rm{g}}{M_{\rm in}} \right)^5\,.  
}
Since the number density of non-thermally produced RHNs is inversely proportional to the initial PBH mass, the production of the non-thermal RHNs is consequently suppressed as the initial PBH mass increases.  

\begin{figure}[ht]
\centering
\includegraphics[width=0.6\textwidth]{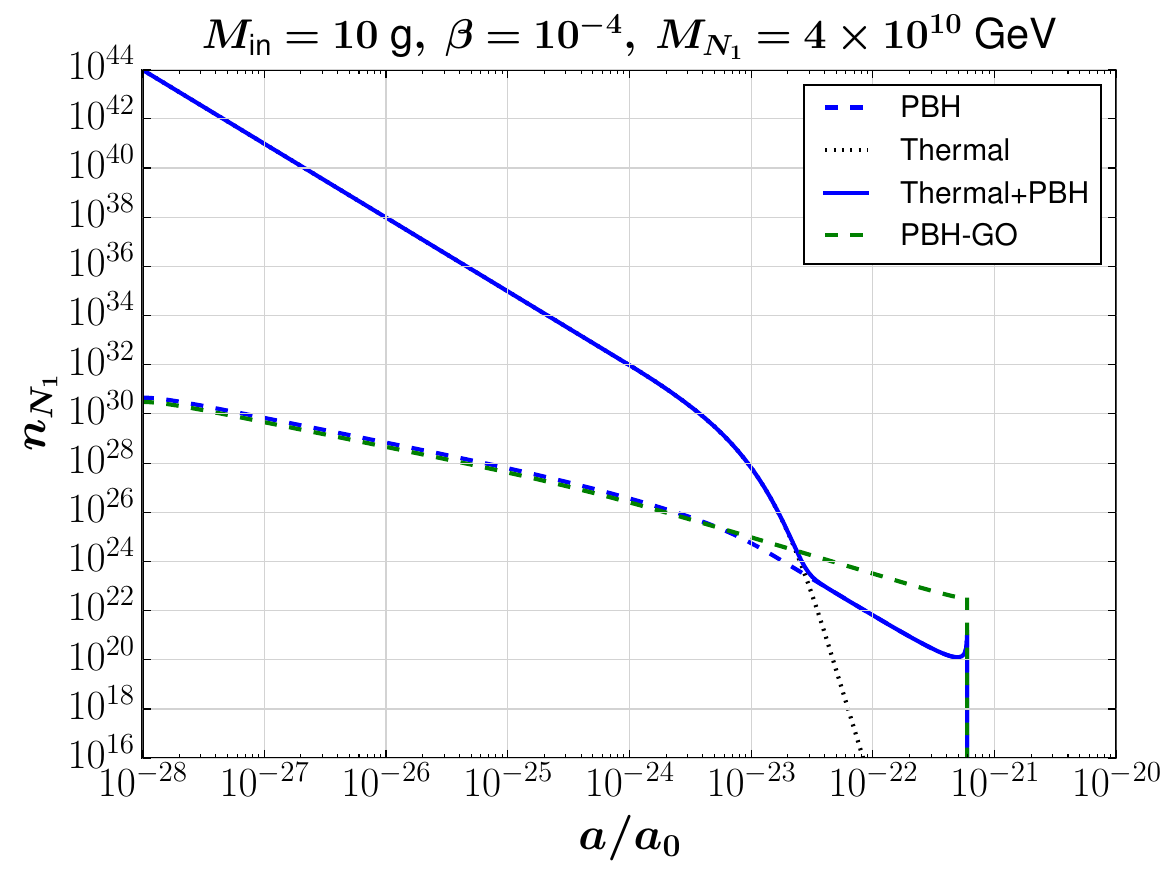}
\caption{Evolution of the RHN number density as a function of the scale factor. Here, we take $M_{\rm in} = 10 \ \rm{g}$, $M_{N_1} = 4 \times 10^{10} \ \rm{GeV}$, and $\beta=10^{-4}$. The black dotted and the blue dashed lines represent the RHN number densities produced via thermal processes and PBH evaporation, respectively, while the blue solid line displays the total contribution. The green dashed line corresponds to the GO limit, which implies the analytic estimation of the non-thermally produced RHNs from PBH evaporation given in Eq.~(\ref{eq:nRHN_GO}).
}
\label{fig:numDen}
\end{figure}

In Fig.~\ref{fig:numDen}, we compare the different contributions to the production of RHNs: (i) thermal production only in the absence of PBHs (black dotted  line), (ii) non-thermal production solely from PBH evaporation (blue dashed  line), (iii) non-thermal production from PBH evaporation in the GO limit (green dashed line), and (iv) the combined thermal and non-thermal contributions (blue solid line). The excellent agreement between the green dashed line, which represents the numerical solution obtained in the GO limit, and the analytic estimation in Eq.~(\ref{eq:nRHN_GO}) validates our numerical implementation of PBH-induced RHN production. As shown by the evolution of the total number density of RHNs, around $a/a_0 \sim 3 \times 10^{-23}$ the production of non-thermal RHN from PBH overtakes the thermal contribution due to the entropy injection into the thermal bath during PBH evaporation. Nevertheless, as the Universe continues to evolve, the thermal contribution eventually dominates the total RHN number density evolution. Furthermore, increasing $M_{\rm in}$ suppresses the non-thermal production of RHNs, in agreement with the analytic scaling shown in Eq.~(\ref{eq:nRHN_GO}).

\subsection{Leptogenesis}
\label{sec:lepto_sub}
In order to track the lepton asymmetry, we need to account for the comoving number density of $N_1$ produced thermally from the plasma, in addition to the non-thermal production via PBH evaporation described by Eq.~(\ref{eq:boltzeqN}). 
The Boltzmann equations governing the cosmological evolution of the thermally produced RHN number density and  the $({B-L})$ asymmetry generated through the decay of the RHNs are written as~\cite{Perez-Gonzalez:2020vnz, Barman:2021ost}
\begin{eqnarray}\label{eq:BE}
    \frac{d \tilde{n}_{N_1}^{\rm TH}}{d \ln(a)} &=&  - \frac{{\Gamma_{N_1}^{\rm TH}}}{{H}}  (\tilde{n}_{N_1}^{\rm TH} - \tilde{n}_{N_1}^{\rm eq})\,, \\
        \frac{d \tilde{n}_{B-L}}{d \ln (a)} &=& 
        \varepsilon_1 \left[(\tilde{n}_{N_1}^{\rm 
        TH}-\tilde{n}_{N_1}^{\rm eq}) \frac{{\Gamma_{N_1}^{\rm TH}}}{{H}} + \tilde{n}_{N_1}^{\rm BH} \frac{{\Gamma_{N_1}^{\rm BH}}}{{H}}\right] - {\rm{Br}}_{N_1\rightarrow l} \frac{W_{1}}{{H}} \tilde{n}_{B-L}\,,\label{eq:BE_B-L}
\end{eqnarray}
where $\Gamma_{N_1}^{\rm TH}= \Gamma_{N_1} {K_1(M_{N_1}/T)}/{K_2(M_{N_1}/T)}$ is the thermally averaged decay rate and $n_{N_1}^{\rm eq}$ is the equilibrium number density of the RHNs as given by 
\dis{
n_{N_1}^{\rm eq} = \frac{g_{N_1}}{2 \pi^2} M_{N_1}^2 T K_2(M_{N_1}/T)\,.
}
Here, $\varepsilon_1$ is the CP asymmetry parameter generated by $N_1$, ${\rm{Br}}_{N_1\rightarrow l}$ denotes the branching ratio of RHN into leptons, ${\rm{Br}}_{N_1\rightarrow l}=1$, and $W_1$ represents the washout term due to inverse decays 
\dis{
W_1 = \frac{1}{4} \Gamma_{N_1}^{\rm TH} K_2(M_{N_1}/T) \left(\frac{M_{N_1}}{T}\right)^2\,.
}

{We work in the single-flavour approximation and retain only inverse decays in the washout. $\Delta L=2$ scatterings are neglected, which is justified for the RHN masses considered here, $M_{N_1}\lesssim10^{12}$ GeV, at the temperatures at which the asymmetry is generated.}
Considering the interference of tree level diagram with one-loop diagrams the $CP$ asymmetry parameter is written as~\cite{Davidson:2002qv}
\begin{equation}
    \varepsilon_1 = -\frac{1}{8\pi}\sum_{j=2,3} \frac{Im[(Y^\dagger Y)_{1j}^2]}{(Y^\dagger Y)_{11}}\bigg[f_v\bigg(\frac{M_j^2}{M_{N_1}^2}\bigg)+f_s\bigg(\frac{M_j^2}{M_{N_1}^2}\bigg)\bigg]
\end{equation}
where the functions $f_v(x)$ and $f_s(x)$ arise from the one-loop vertex and self-energy corrections, respectively, and are given by
\begin{equation}
\begin{split}
    &f_s(x)=\frac{\sqrt{x}}{1-x}\,,\\
    &f_v(x)= \sqrt{x}\bigg[1-(1+x)\ln\bigg(\frac{1+x}{x}\bigg)\bigg]\,.
\end{split}
\end{equation}
Using the Davidson--Ibarra bound, one can find the simple expression for the $CP$ asymmetry given by 
\begin{equation}
\varepsilon_1
\simeq
\frac{3}{16\pi}
\frac{m_\nu M_{N_1} }{v^2} \simeq 9.85 \times 10^{-6} \left(\frac{M_{N_1}}{10^{11} \ \rm{GeV}}\right)\,,   
\end{equation}
where we take $m_\nu = 0.05 \ \rm{eV}$ and $\upsilon=174 \ \rm{GeV}$ in the last step{, and where we work throughout in the hierarchical, non-resonant regime $M_{N_1}\ll M_{N_{2,3}}$}.

\begin{figure}[ht]
\centering
\includegraphics[width=0.49\textwidth]{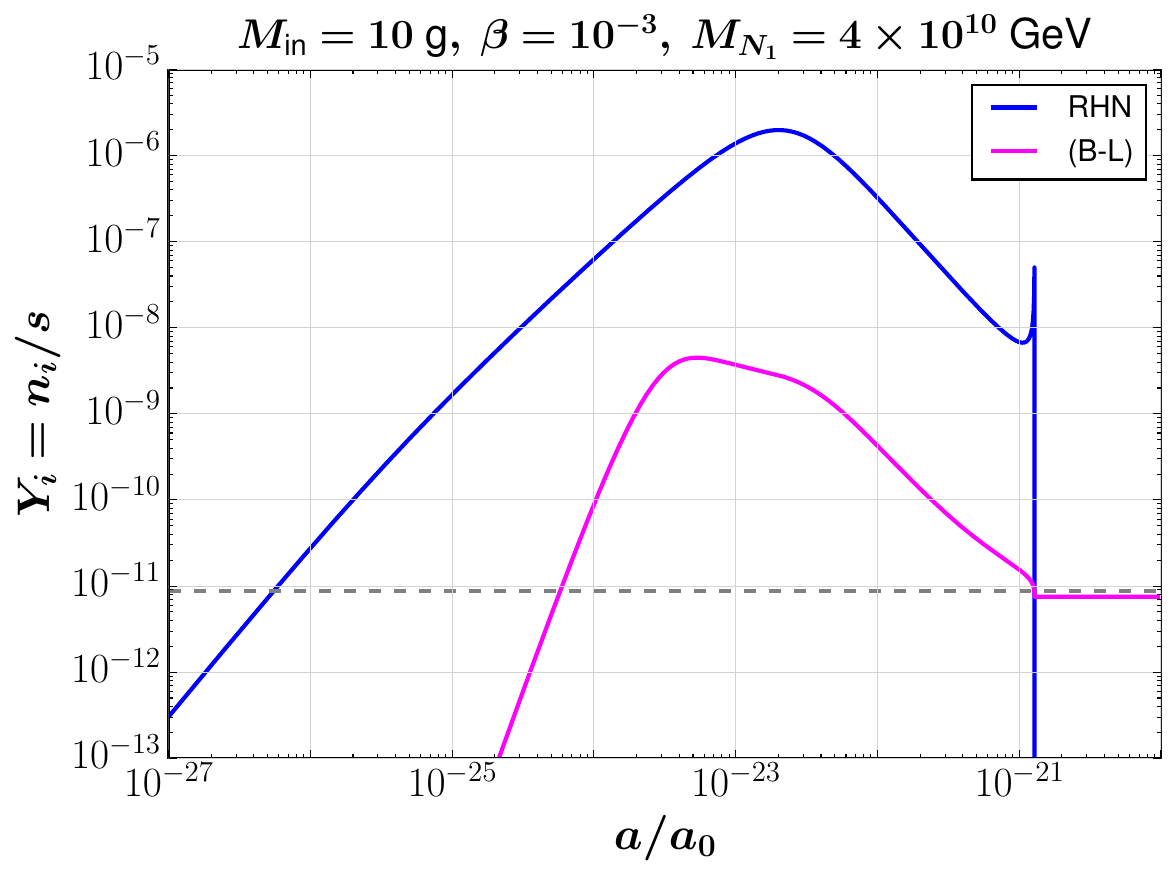}
\hfill
\includegraphics[width=0.49\textwidth]{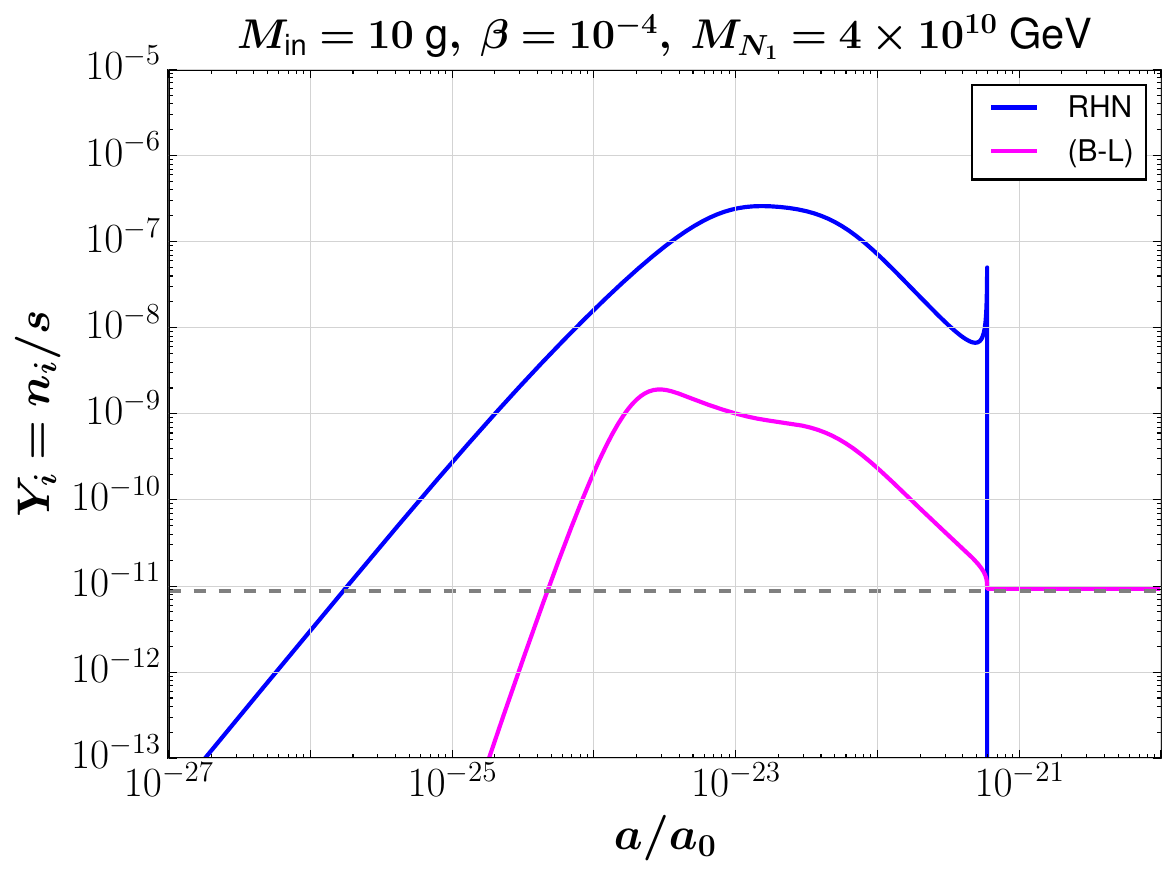}
\hfill
\includegraphics[width=0.49\textwidth]{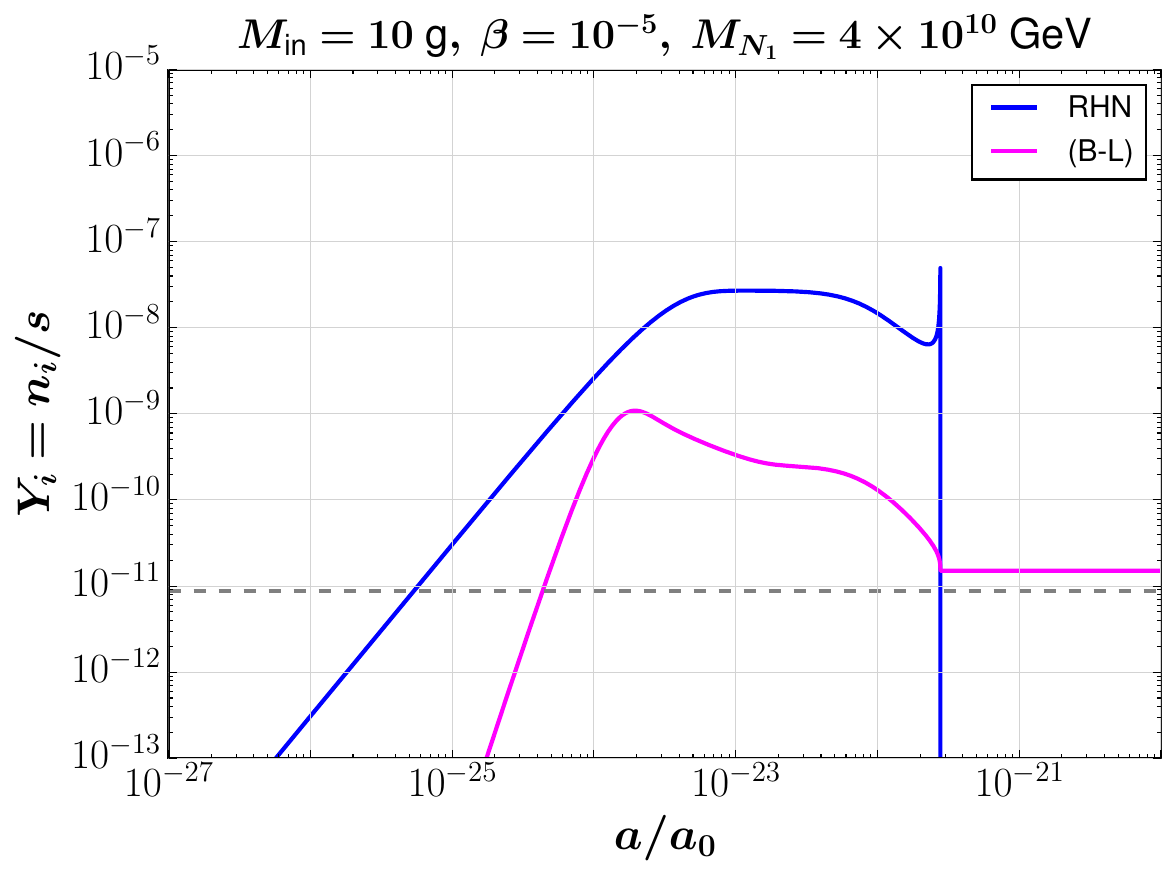}
\hfill
\includegraphics[width=0.49\textwidth]{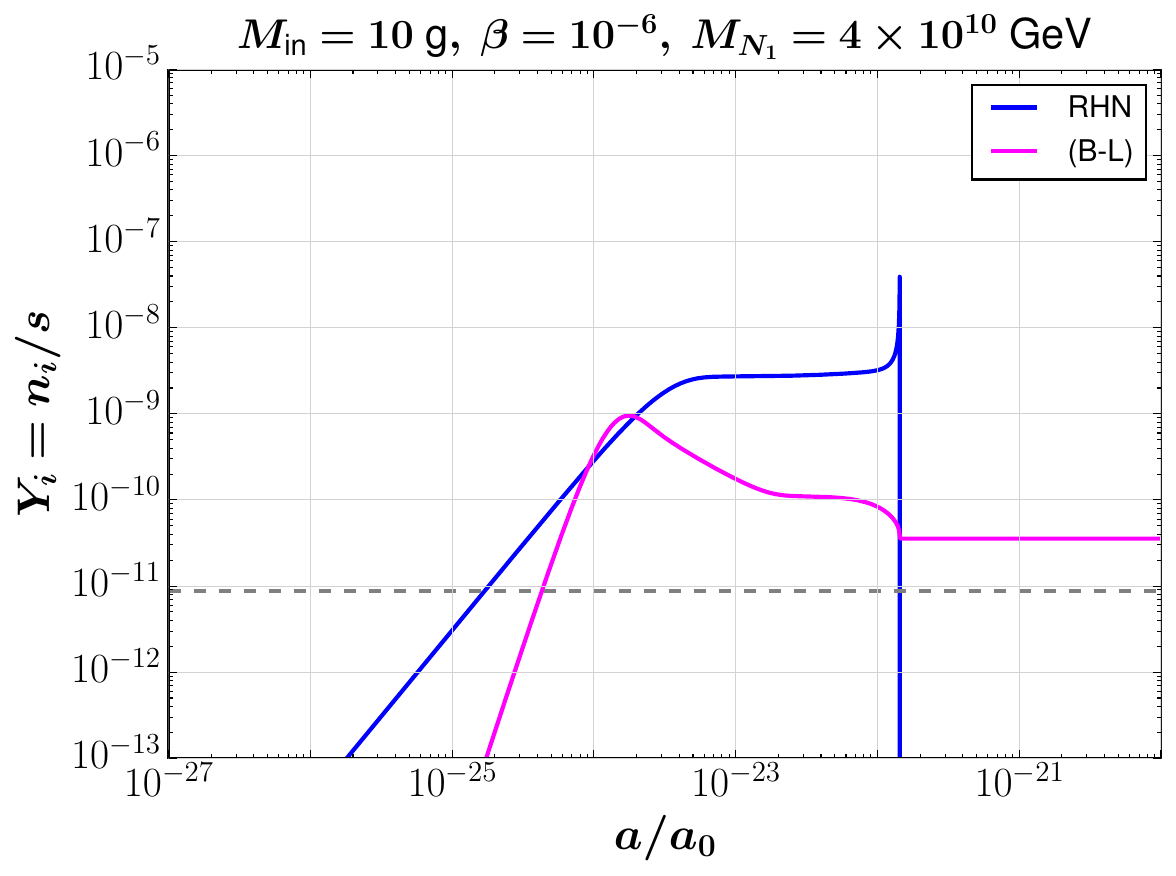}
\caption{Evolution of the yields of the non-thermal RHN, $N_1$, produced from PBH evaporation along with the ($B-L$) asymmetry as a function of the scale factor for the different $\beta$ parameters.
 The horizontal dashed gray line represents the observed ($B-L$) asymmetry with respect to $Y_B^{\rm obs}\simeq8.7\times10^{-11}$.
}
\label{fig:Y_(B-L)}
\end{figure}


{The final $(B-L)$ asymmetry is converted into a baryon asymmetry by the electroweak sphalerons according to
\begin{equation}\label{eq:YB_conversion}
    Y_B \equiv \frac{n_B}{s} = C_{B\to L}\, Y^{\rm f}_{B-L}\,,
    \qquad
    C_{B\to L}=\frac{28}{79}=0.35443\,,
\end{equation}
where $Y^{\rm f}_{B-L}=\tilde{n}_{B-L}/(s\,a^3)$ is evaluated once PBH evaporation is complete and the asymmetry has saturated. We emphasise that no separate dilution factor is required in Eq.~(\ref{eq:YB_conversion}) as the entropy density $s$ is taken from the full numerical solution of Eq.~(\ref{eq:Tem_Boltz}), and therefore already includes the entropy injected by the evaporating black holes as well as the standard reduction in $g_{*,s}$ between leptogenesis and recombination. A fixed dilution factor of the form $g_*(T_{\rm lep})/g_{*,s}(T_0)$, appropriate to a standard thermal history, would not capture the PBH entropy release, which is the dominant dilution effect in the present scenario. For reference, the corresponding baryon-to-photon ratio today is $\eta_B = 7.04\,Y_B$, and the observed value $Y_B^{\rm obs}\simeq8.7\times10^{-11}$ requires $Y^{\rm f}_{B-L}\simeq2.5\times10^{-10}$. }

In Fig.~\ref{fig:Y_(B-L)}, we show the evolution of the yields of non-thermally produced RHNs from PBH evaporation together with the $(B-L)$ asymmetry as a function of the scale factor for four different initial PBH abundances, $\beta$. For $M_{\rm in} = 10 \ \rm{g}$ case, the minimum initial PBH abundance required for the PBH dominated epoch is $\beta_{\rm min} \simeq 6.8 \times 10^{-7} \ \rm{g}$. As a consequence, for $\beta=10^{-6}$ (lower right panel), the PBH dominated era is shorter than in the cases with larger initial abundances, leading to reduced production of non-thermal RHNs. 
The figure clearly shows that the dependence of the final $(B-L)$ asymmetry on $\beta$ is relatively weak once the Universe undergoes the PBH dominated epoch. 

The $(B-L)$ asymmetry initially grows due to the thermal contribution of RHNs and subsequently diminishes due to the washout effect induced by the inverse decays. During the PBH evaporation, the entropy injection into the thermal bath causes an additional dilution of the asymmetry. After the PBHs have completely evaporated, the $(B-L)$ asymmetry saturates reaching a constant value. The dilution effect due to the entropy injection becomes more prominent for massive PBHs as they evaporate over a longer duration.

In this work, we consider the scenario in which the lepton asymmetry generated by decay of RHNs, including both thermal and non-thermal contributions described by Eq.~(\ref{eq:BE_B-L}),
is converted into observed baryon asymmetry through the $(B-L)$-violating EW sphaleron transitions. In the limiting case where the lepton asymmetry is generated solely by the non-thermal RHNs produced from PBH evaporation, one needs to require that PBH evaporation should be completed before the EW sphalerons freeze out in order for the generated lepton asymmetry to be efficiently converted into the baryon asymmetry. This requires the PBH evaporation temperature during the PBH dominated epoch to satisfy $T_{\rm ev}|_{\rm MD} \gtrsim T_{\rm EW}\simeq 130 \ \rm{GeV}$. Therefore, using Eq.~(\ref{eq:evapTemperature}), the corresponding upper bound on the initial mass of PBH is found as 
\dis{\label{eq:Min_EWPT}
M_{\rm in } \lesssim 4.2 \times 10^5 \ \rm{g}.
}

\begin{figure}[ht]
\centering
\includegraphics[width=0.49\textwidth]{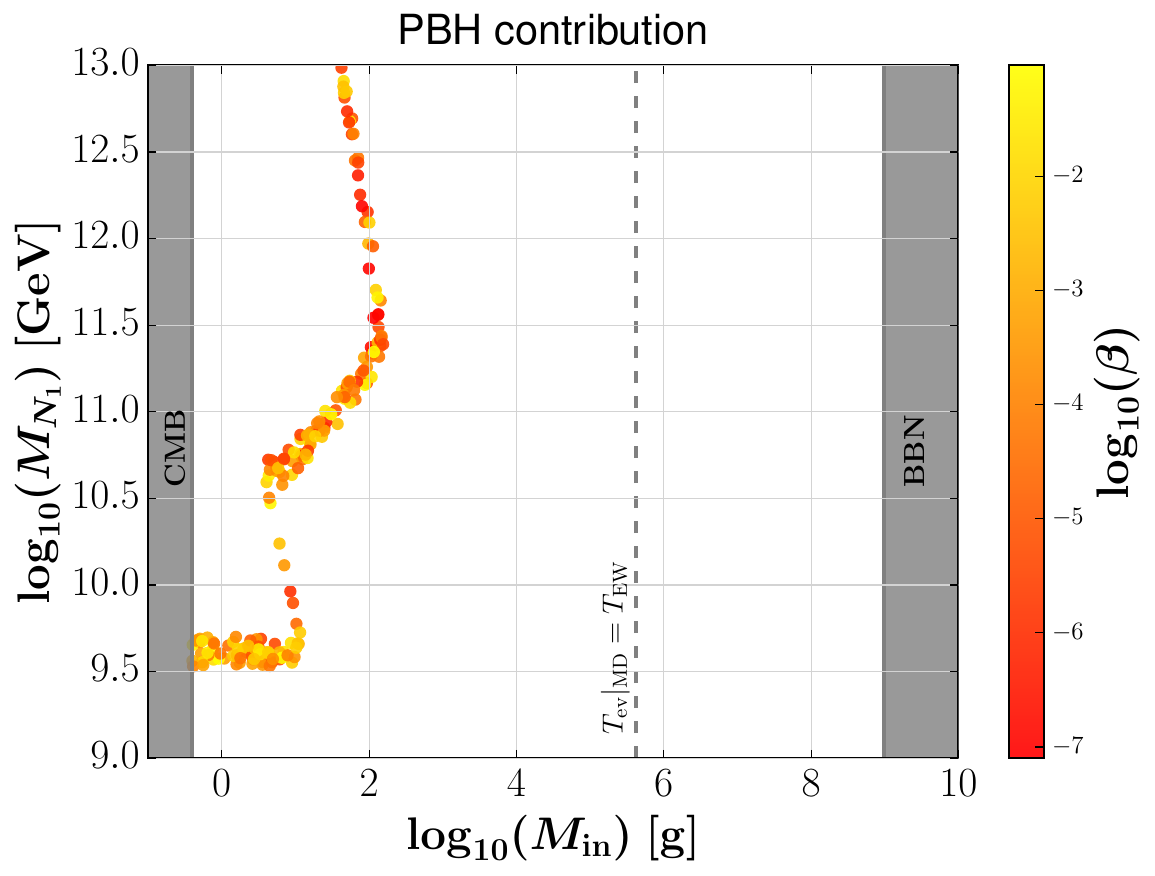}
\hfill
\includegraphics[width=0.49\textwidth]{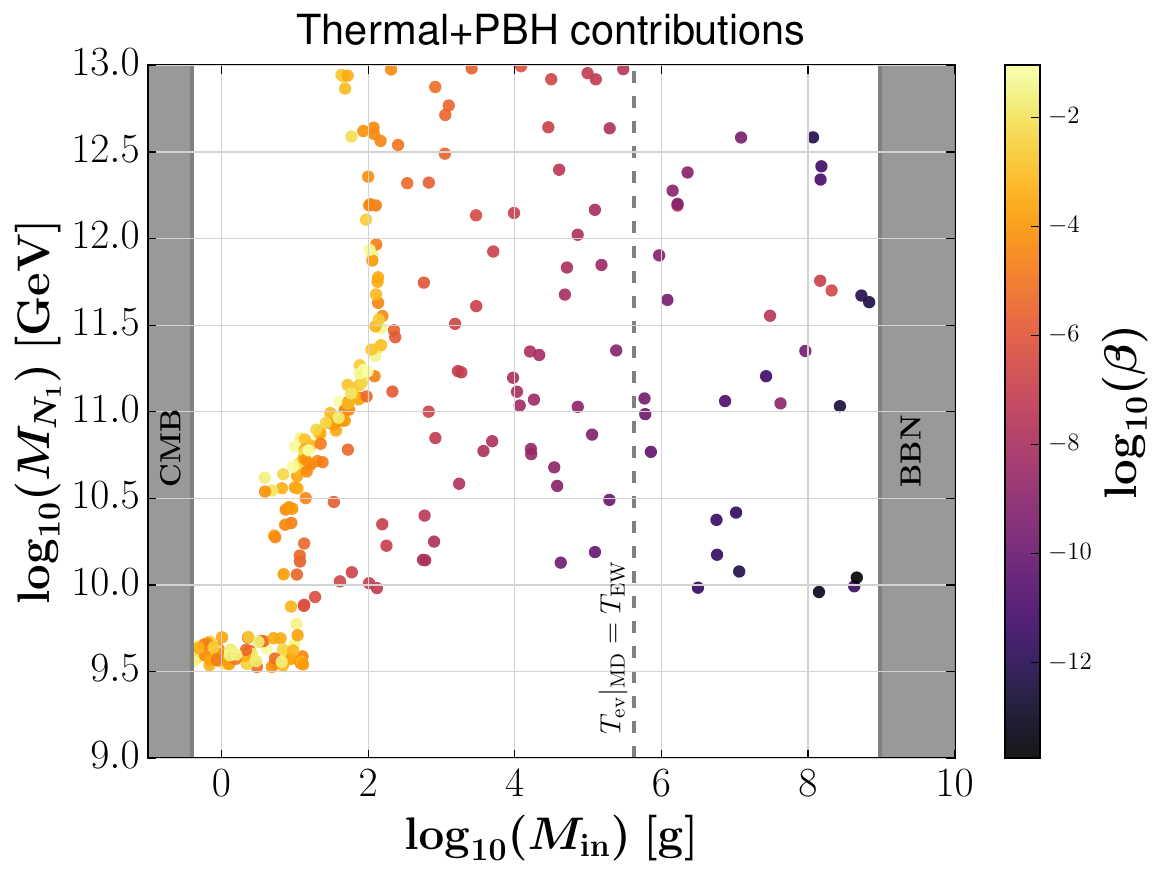}
\caption{Parameter scan for the observed baryon asymmetry in the $\log_{10}(M_{N_1})$--$\log_{10} (M_{\rm in})$ plane taking into account only PBH contribution (left) and both thermal and PBH contributions (right). The upper and lower bounds on the initial PBH mass~\cite{Carr:2020gox} are imposed by the BBN and CMB, respectively. The gray dashed line indicates the sphaleron transition, corresponding to the upper bound on the initial PBH mass derived in Eq.~(\ref{eq:Min_EWPT}). {As explained in the text, this bound applies only to the PBH-only scenario of the left panel and is shown in the right panel for reference only}. }
\label{fig:mN_vs_Min_scan}
\end{figure}

In Fig.~\ref{fig:mN_vs_Min_scan}, we perform a parameter scan to identify the region consistent with the observed baryon asymmetry by randomly varying the initial PBH mass and the RHN mass over the ranges $M_{\rm in} \in [0.4, \ 10^9] \ \rm{g}$ and $M_{N_1} \in [10^5, \ 10^{13}] \ \rm{GeV}$, respectively. For each parameter point, the initial PBH abundance, $\beta$, is chosen to satisfy the condition for the PBH dominated epoch. The allowed range of the initial mass of PBH, $0.4 \ \text{g} \lesssim M_\text{in} \lesssim 9.7 \times 10^8 \ \text{g}$~\cite{Carr:2020gox}, is indicated by gray shaded regions. As a reference, the upper bound on $M_{\rm{in}}$ derived from Eq.~(\ref{eq:Min_EWPT}), corresponding to the requirement that PBH evaporation be completed before the EW sphalerons freeze out, is shown by the gray dashed line. The left panel of Fig.~\ref{fig:mN_vs_Min_scan} shows the parameter space capable of generating the observed baryon asymmetry when only the PBH contribution to leptogenesis is taken into account, whereas the right panel includes both the thermal and PBH contributions. One can see that smaller initial PBH masses, $0.4 \lesssim M_{\rm in} \lesssim 10^2 \ \rm{g}$\footnote{Ref.~\cite{Barman:2021ost} found a narrower allowed range for the initial PBH mass, $0.5 \lesssim M_{\rm in} \lesssim 10 \ \rm{g}$, based on the requirement of reproducing the observed baryon asymmetry from PBH evaporation.}, are more efficient to generate the baryon asymmetry in the PBH only scenario than in the case where both production mechanisms are included. 
Moreover, for approximately $M_{\rm in} \gtrsim 10^2 \ \rm{g}$, leptogenesis is predominantly driven by the thermal production of RHNs, while the contribution from PBH evaporation becomes subdominant.

\begin{figure}[ht]
\centering
\includegraphics[width=0.49\textwidth]{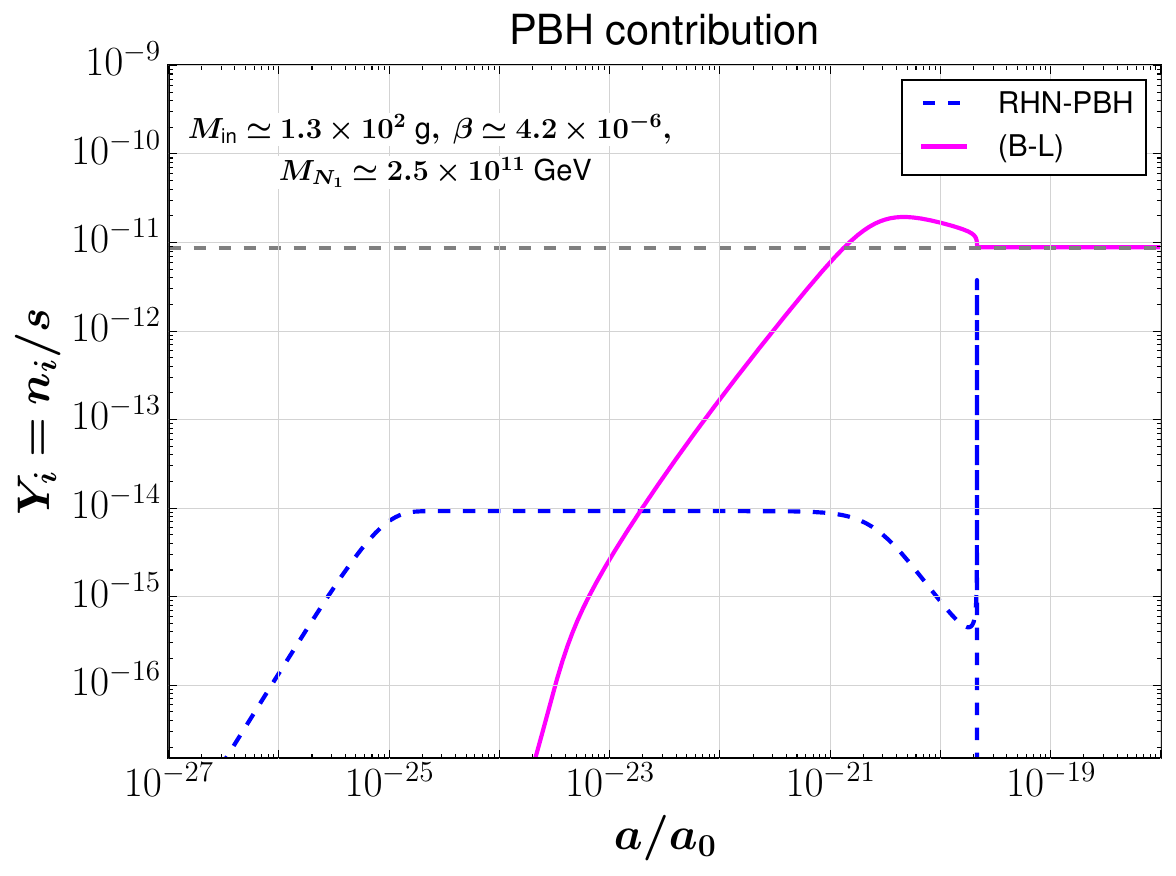}
\hfill
\includegraphics[width=0.49\textwidth]{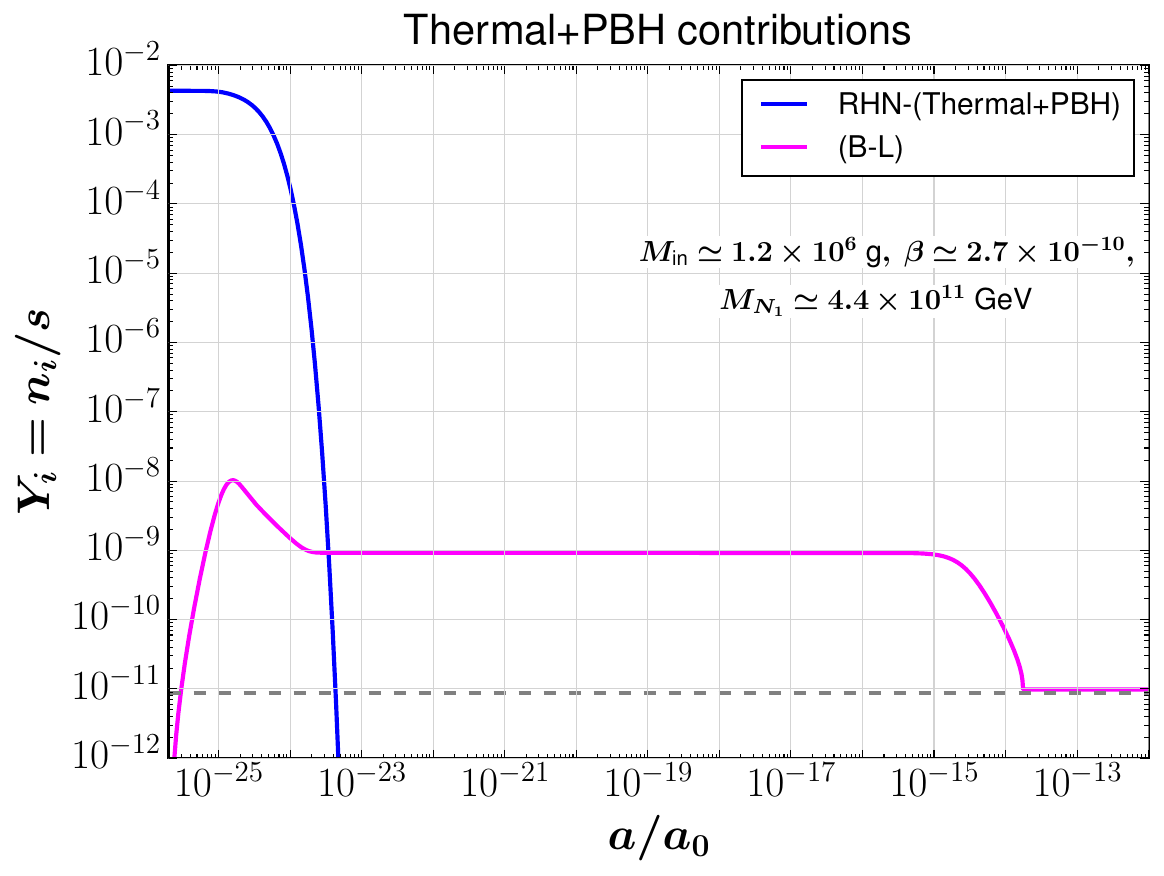}
\caption{Evolution of the yields considering two different scenarios: leptogenesis driven solely by PBH produced RHNs (left) and by the combined thermal and PBH produced RHNs (right). The benchmarks values shown in the plots are chosen from the parameter scan presented in Fig.~\ref{fig:mN_vs_Min_scan}.}
\label{fig:Y_benchmark}
\end{figure}

In Fig.~\ref{fig:Y_benchmark}, we show the evolution of the relevant yields for two different scenarios: considering only the PBH contribution (left) and the combined thermal and PBH contributions (right), respectively. The corresponding benchmark parameter values are selected from the parameter scan presented in Fig.~\ref{fig:mN_vs_Min_scan}{, and correspond to the benchmark points BP1 and BP2 of Table~\ref{tab:benchmarks}}. Furthermore, in the combined scenario, the chosen benchmark value of $M_{\rm in} \sim 10^6 \ \rm{g}$ simultaneously reproduces the correct dark matter relic abundance through Planck scale PBH remnants. The corresponding initial PBH mass is consistent with the value obtained in Eq.~(\ref{eq:Min_PlanckRelic}).  

{It is important to be precise about the scope of Eq.~(\ref{eq:Min_EWPT}). It applies only to the limiting case in which the entire $B-L$ asymmetry is carried by right-handed neutrinos emitted by the black holes, and it is therefore shown as a constraint only in the left panel of Fig.~\ref{fig:mN_vs_Min_scan}. When the thermal channel contributes, the asymmetry is generated at $T\sim M_{N_1}\gg T_{\rm EW}$, long before evaporation, and is reprocessed by the sphalerons in the standard way at $T_{\rm EW}\simeq130$ GeV. Evaporation at a later time, $T_{\rm ev}|_{\rm MD}<T_{\rm EW}$, then no longer destroys the asymmetry but merely dilutes it: the baryon number per comoving volume is conserved once the sphalerons have frozen out, while the entropy is increased by the black hole decay. This dilution is fully included in the numerical solution through Eq.~(\ref{eq:Tem_Boltz}) and is visible as the drop in $Y_{B-L}$ near $a/a_0\sim10^{-14}$ in the right panel of Fig.~\ref{fig:Y_benchmark}. We note in passing that the sphaleron freeze-out temperature is set by the electroweak crossover, through the exponential suppression of the sphaleron rate once the Higgs acquires its vacuum expectation value, rather than by a competition between the sphaleron rate and the expansion rate; the modified expansion history during the PBH-dominated epoch therefore does not shift $T_{\rm EW}$ appreciably.}

\section{Dark Matter from Planck Scale Remnants of PBH }
\label{sec:dm}
While the semiclassical Hawking evaporation predicts a complete mass loss, quantum gravitational backreaction is expected to become significant as the black hole mass approaches the Planck scale {$M_{\rm Pl}=(8\pi)^{1/2}M_{p} \simeq 1.22\times10^{19} \ {\rm GeV}\simeq 2.17 \times 10^{-5}$ g}. Theories such as Loop Quantum Gravity suggest that the event horizon may be replaced by a quantum-corrected structure, preventing the singularity and halting evaporation. As a result, a stable, non-radiating remnant may form with mass $m_{\rm rem} = r M_{\rm Pl}$, where $r \ge 1$ is an order-one coefficient{, whose precise value is not predicted by any of the candidate frameworks and which we therefore retain as a free parameter throughout}. Such remnants are natural candidates for cold dark matter~\cite{Domenech:2023mqk, Trivedi:2025vry, Lehmann:2019zgt, Ong:2024dnr, Sasaki:2025zao}. They are compact, massive compared to elementary particles, and interact predominantly through gravity, rendering them effectively invisible to both direct detection and astrophysical searches. To assess whether these remnants can account for the observed dark matter abundance today, we must calculate their relic density within our PBH-assisted leptogenesis scenario. 

{The abundance at the moment of evaporation follows from a simple counting argument. Each evaporating black hole leaves behind exactly one remnant, so $n_{\rm rem}(t_{\rm ev}) = n_{\rm BH}(t_{\rm ev})$, while the energy released into the plasma is $(M_{\rm in}-m_{\rm rem})n_{\rm BH}(t_{\rm ev})$ per unit volume. If the black holes dominate the energy density before evaporating, essentially all of the ambient energy at $t_{\rm ev}$ originates from the black holes themselves, and therefore}
the fraction of the relic density of PBH remnants at the moment of evaporation is given by
\dis{\label{eq:Omega_rem}
\Omega_{\rm rem} (t_{\rm ev}) = \frac{\rho_{\rm rem}(t_{\rm ev})}{\rho_{\rm tot}(t_{\rm ev})}\,{= \frac{m_{\rm rem}\,n_{\rm BH}(t_{\rm ev})}{M_{\rm in}\,n_{\rm BH}(t_{\rm ev})}
= \frac{m_{\rm rem}}{M_{\rm in}}
= r\,\frac{M_{\rm Pl}}{M_{\rm in}}
\simeq 2.17\times10^{-5}\,r\left(\frac{1~{\rm g}}{M_{\rm in}}\right)\,,}}
{up to corrections of order $m_{\rm rem}/M_{\rm in}$. The remnant fraction at evaporation is thus nothing other than the fraction of the initial PBH mass that survives, and in particular it is entirely independent of the initial abundance $\beta$. This is the central simplification afforded by the PBH-dominated history assumed throughout, and it is the reason we restricted attention to that case in Sec.~\ref{sec:PS}.}

{It is useful to verify Eq.~(\ref{eq:Omega_rem}) against the explicit expansion history. Writing $\rho_{\rm rem}(t_{\rm ev}) = m_{\rm rem} n_{\rm BH}(t_f)(a_f/a_{\rm ev})^3$ with $n_{\rm BH}(t_f)$ from Eq.~(\ref{eq:nBHin}), and $\rho_{\rm tot} = 3M_p^2H^2$ at each epoch, one has
\dis{\label{eq:Om_rem}
\Omega_{\rm rem}(t_{\rm ev}) = \frac{\beta\, m_{\rm rem}}{M_{\rm in}}
\left(\frac{H^2_f}{H^2_{\rm ev}}\right)\left(\frac{a_f}{a_{\rm ev}}\right)^3 .}}
{During a PBH-dominated epoch $3M_p^2H^2 \simeq \rho_{\rm BH} = \beta\,\rho_{\rm tot}(t_f)(a_f/a)^3$, so that
\begin{equation}
\beta\left(\frac{H^2_{f}}{H^2_{\rm ev}}\right)\left(\frac{a_f}{a_{\rm ev}}\right)^3 = 1 \,,
\end{equation}
and Eq.~(\ref{eq:Om_rem}) reduces to Eq.~(\ref{eq:Omega_rem}), as it must. 
}

Defining the fraction of the PBH remnants in the total DM density today as $f_{\rm rem} \equiv \frac{\rho_{\rm rem,0}}{\rho_{\rm DM,0}}$, and noting that $\rho_{\rm rem} \propto a^{-3}$, Eq.~(\ref{eq:Omega_rem}) can equivalently be written as 
\dis{
\Omega_{\rm rem} (t_{\rm ev}) = f_{\rm rem} \frac{\rho_{\rm DM, 0}}{\rho_{\rm tot}(t_{\rm ev})} \left(\frac{a_0}{a_{\rm ev}}\right)^3\,.
}
The ratio of the scale factor at evaporation to the present value is approximately:
\begin{equation}\label{eq:aevtoa0}
    \frac{a_{\rm ev}}{a_0} \simeq \frac{T_0}{T_{\rm ev}} \left( \frac{g_{*,s}(T_0)}{g_{*,s}(T_{\rm ev})} \right)^{1/3}
    \simeq 2.3 \times 10^{-24} \left( \frac{M_{\rm in}}{1\, \mathrm{g}} \right)^{3/2}\,,
\end{equation}
where we have used Eq.~(\ref{eq:evapTemperature}).
Taking $\rho_{\rm tot}(t_{\rm ev}) = \frac{\pi^2}{30} g_*(T_{\rm ev}) T_{\rm ev}^4$ and $\Omega_{\rm DM, 0} h^2=0.12$, $\Omega_{\rm rem} (t_{\rm ev})$ can be simplified as 
\dis{
\begin{aligned}\label{eq:OmRem_wrt_fRem}
\Omega_{\rm rem} (t_{\rm ev}) &= f_{\rm rem} \rho_{\rm cr,0} \Omega_{\rm DM, 0}
\frac{30}{\pi^2 g_*(T_{\rm ev}) T_{\rm ev} T_0^3}
\left( \frac{g_{*,s}(T_{\rm ev})}{g_{*,s}(T_0)} \right) \\
&\simeq 5.8 \times 10^{-10} f_{\rm rem}
\left(\frac{\rm GeV}{T_{\rm ev}}\right)
\simeq 1.6 \times 10^{-20} f_{\rm rem}
\left(\frac{M_\text{in}}{1~\text{g}}\right)^{3/2}\,,
\end{aligned}
}
where we have used $g_{*,s}(T_0) \simeq 3.91$, $g_{*,s}(T_{\rm ev}) \simeq g_{*}(T_{\rm ev})$, $T_0 \simeq 2.34 \times 10^{-13} \ \rm{GeV}$ and $\rho_{\rm cr,0} \simeq 8 h^2 \times 10^{-47} \ \rm{GeV}^4$. 
Combining Eqs.~(\ref{eq:Omega_rem}) and (\ref{eq:OmRem_wrt_fRem}), the initial PBH mass is determined as a function of $r/f_{\rm rem}$
{\dis{\label{eq:Min_PlanckRelic}
\left(\frac{M_\text{in}}{1~\text{g}}\right) \simeq {1.1} \times 10^6 \left(\frac{r}{f_{\rm rem}}\right)^{2/5}\,,}
or equivalently
\begin{equation}\label{eq:frem}
f_{\rm rem} \simeq r\left(\frac{1.1\times10^{6}~{\rm g}}{M_{\rm in}}\right)^{5/2}.
\end{equation}}
Assuming that PBH remnants constitute the total DM density today, i.e., $f_{\rm rem}=1$, with $r=1$, the required initial PBH mass in the PBH dominated scenario becomes 
{$M_{\rm in} \simeq 1.1 \times 10^{6} \ \rm{g}$}. 
{This is within ten percent of the benchmark BP2 of Table~\ref{tab:benchmarks}, $M_{\rm in}\simeq1.2\times10^{6}$ g, which was chosen from the leptogenesis scan of Fig.~\ref{fig:mN_vs_Min_scan} and corresponds to $f_{\rm rem}\simeq0.8\,r$.}

{The coefficient $r$ is not predicted by any of the frameworks in which
remnants arise. Hence it can be tuned according to the initial mass to achieve the correct DM relic density. Setting $f_{\rm rem}=1$
in Eq.~(\ref{eq:frem}) gives
\begin{equation}\label{eq:r_of_Min}
r = \left(\frac{M_{\rm in}}{1.1\times10^{6}~{\rm g}}\right)^{5/2}\,.
\end{equation}
Thus remnants can account for the entire dark matter abundance, for any
initial mass above $\simeq1.1\times10^{6}$ g by a suitable choice of
$r\ge1$. For BP2, this requires only $r\simeq 1.2$. Here, it is worth mentioning that this freedom is one-sided as for lighter black holes, it would demand
$r<1$ which is not allowed and hence the remnants overclose the Universe. Because the remnant fraction falls as the five-halves power of the
initial mass, a reduction of the mass by only ten percent is enough to
overclose the Universe for any permitted $r$, while an increase of the
same size leaves a remnant abundance too small to account for the dark
matter unless $r$ departs appreciably from unity. Thus, the naturalness of an order-unity coefficient bounds the initial mass from above as well.}


\section{Gravitational Waves}\label{sec:GW}
\label{sec:gw}

Because Pati--Salam symmetry breaking occurs prior to inflation, any stochastic gravitational wave background generated by the associated first-order phase transition \cite{Caprini:2015zlo,Hindmarsh:2020hop} is exponentially diluted alongside the magnetic monopoles, its energy density redshifting as $e^{-4N}$ once the relevant modes are pushed outside the horizon. The same conclusion applies to the domain walls that accompany D-parity breaking \cite{Kibble:1976sj}. Topological defects of the one type that would survive, namely cosmic strings, do not arise here: for the breaking chain of Eq.~(\ref{eq:chain}) the relevant homotopy group is $\pi_1(G_{\rm PS}/H)=\pi_0(H)$, which vanishes because the unbroken group $H=SU(3)_C\times U(1)_Y$ is connected. The gravitational wave phenomenology of this framework is therefore carried not by the gauge sector but by the primordial black hole population itself, through three distinct second-order channels: the scalar-induced background associated with the curvature perturbations that form the black holes, the tensor modes sourced by the Poisson fluctuations of the black hole distribution, and the direct emission of gravitons during Hawking evaporation.

The first channel is the scalar-induced background generated at second order by the enhanced curvature perturbations required for PBH formation \cite{Ananda:2006af,Baumann:2007zm,Kohri:2018awv,Domenech:2021ztg}. An amplitude $\mathcal{P}_\zeta(k_f)\sim\mathcal{O}(10^{-2})$ of the kind discussed in Sec.~\ref{sec:inflation} generically sources a tensor spectrum of comparable prominence, with $\Omega_{\rm GW}\sim\mathcal{P}_\zeta^{2}$ near the peak. Since the peak tracks the formation scale $k_f$ of Eq.~(\ref{eq:k_f}), and $k_f\propto M_{\rm in}^{-1/2}$, the signal associated with the gram-scale to $10^{6}$ g black holes considered here redshifts to frequencies of order $10^{4}$--$10^{8}$ Hz, with the entropy released during evaporation both lowering the peak frequency and suppressing the amplitude. This lies far above the band of any current or planned interferometer, and its detection would require the high-frequency detector concepts currently under development \cite{Aggarwal:2020olq}, the present-day peak frequency being \cite{Saito:2008jc, Domenech:2020ssp}
\begin{equation}
f_{\rm peak} \simeq 1.2\times10^{8}\,{\rm Hz}
\left(\frac{1\,{\rm g}}{M_{\rm in}}\right)^{1/2}
\left(\frac{\beta_{\rm min}}{\beta}\right)^{1/3}.
\label{eq:fpeak_SIGW}
\end{equation}

The second channel is more promising observationally. Because primordial black holes form at rare peaks of the density field, their spatial distribution carries Poisson fluctuations, which behave as isocurvature perturbations at formation, are converted into curvature perturbations once the black holes dominate, and source tensor modes at the abrupt end of the matter-dominated epoch \cite{Papanikolaou:2020qtd,Inomata:2020lmk,Domenech:2020ssp}. This background peaks at the comoving scale of the mean black hole separation, corresponding today to $f\simeq1.7\ {\rm kHz}\,(M_{\rm in}/10^{4}\,{\rm g})^{-5/6}$, which for the heavier masses selected by the remnant dark matter condition of Eq.~(\ref{eq:Min_PlanckRelic}) falls in the range of tens of hertz, { for example BP1 gives $f\simeq35$ Hz at $M_{\rm in}=1.2\times10^{6}$ g,} within the reach of the Einstein Telescope \cite{Punturo:2010zz} and Cosmic Explorer \cite{Reitze:2019iox}. Its amplitude scales steeply with the initial abundance, and the requirement that it not spoil the effective number of relativistic species at nucleosynthesis imposes $\beta\lesssim1.1\times10^{-6}(M_{\rm in}/10^{4}\,{\rm g})^{-17/24}$ \cite{Domenech:2020ssp}, a bound satisfied by the benchmarks of Sec.~\ref{sec:leptogenesis}.

Since the baryon asymmetry depends only weakly on $\beta$ once the PBHs dominate, while the remnant abundance is entirely independent of $\beta$, the initial PBH abundance remains a free parameter of the scenario. It is constrained only from below by the PBH domination requirement of Eq.~(\ref{eq:betamin}) and from above by the requirement of successful BBN. The {Poisson-induced} stochastic gravitational wave background is the only observable sensitive to the $\beta$, and the corresponding signal may therefore lie anywhere between the two limits. {This makes it, in principle, a very good discriminator of the scenarios as a detection at the Einstein Telescope or Cosmic Explorer would fix the one remaining free parameter of the framework.}
{A quantitative computation of both gravitational wave spectra, together with the resulting constraints on the $(M_{\rm in},\beta)$ plane and their interplay with the electroweak sphaleron condition of Eq.~(\ref{eq:Min_EWPT}), lies beyond the scope of the present work and is left for a dedicated study.}

In addition to the induced stochastic gravitational wave background, PBHs also generate a gravitational wave signal through the direct emission of  gravitons during Hawking evaporation. This contribution, together with graviton bremsstrahlung from the decay of  massive scalar particles, has been investigated in Refs.~\cite{Choi:2024acs, Choi:2025hqt}. However, the resulting gravitational wave spectrum peaks in the ultra-high frequency regime, well beyond the sensitivity of the current and planned gravitational wave detectors. For example, the peak frequency is determined by the initial mass of PBH and is given by 
\dis{
f \simeq 1.6 \times 10^{13} \ {\rm Hz} \left(\frac{M_{\rm in}}{1 \ \rm g}\right)^{1/2} \,.
}
This scaling is opposite to that of Eq. (\ref{eq:fpeak_SIGW}) because heavier black holes emit softer gravitons, but they evaporate later and the emitted gravitons are therefore less redshifted. As a consequence, increasing the initial PBH mass shifts the peak of the gravitational wave towards higher frequencies.

\begin{figure}[ht]
\centering
\includegraphics[width=0.6\textwidth]{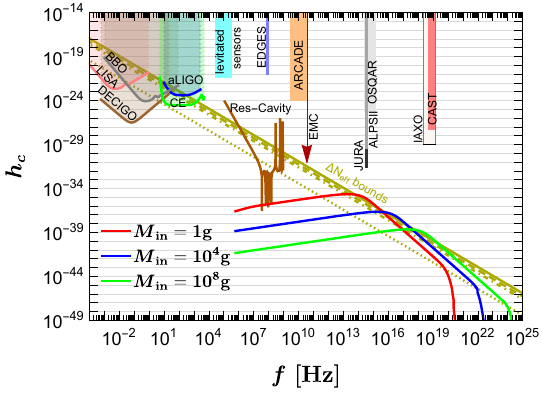}
\caption{Characteristic gravitational wave strain $h_c$ from the direct evaporation of PBHs for three different initial masses, $M_{\rm in}=1, \ 10^{4}, \ 10^{8}\ \rm{g}$. The shaded regions indicate the sensitivities of several current and proposed gravitational wave detectors~\cite{Thrane:2013oya, Reitze:2019iox, Seto:2001qf, KAGRA:2013rdx, Aggarwal:2020olq}. The brown curve shows the projected sensitivity of resonant cavity experiments~\cite{Herman:2022fau}. The dark yellow lines represent constraints on $\Delta N_{\rm eff}$ from the Planck experiment~\cite{Planck:2018vyg} (solid), and the projected sensitivities of CMB-S4~\cite{CMB-S4:2016ple} (dashed), Euclid~\cite{laureijs2011euclid} (dot-dashed), and a proposed cosmic-variance-limited CMB (CMB-CVL) experiment~\cite{Ben-Dayan:2019gll} (dotted).
}
\label{fig:GW_direct}
\end{figure}

For illustration, Fig.~\ref{fig:GW_direct} shows the characteristic gravitational wave strain from direct evaporation of PBH  as a function of frequency $f$ for three representative initial PBH masses, $M_{\rm in} = 1, \ 10^{4}, \ 10^{8} \ \rm g$, displayed by the red, blue and green solid curves, respectively. The shaded regions represent the limits from various proposed GW detectors, including LISA~\cite{Thrane:2013oya}, the Big Bang Observer (BBO)~\cite{Thrane:2013oya}, CE~\cite{Reitze:2019iox}, DECIGO~\cite{Seto:2001qf}, and the advanced LIGO~\cite{KAGRA:2013rdx}. We also display the current and projected sensitivities of several proposed high-frequency gravitational wave detection techniques, including optically levitated sensors, enhanced magnetic conversion (EMC), and the inverse Gertsenshtein effect, as well as the experiments JURA, ALPS II, OSQAR, IAXO, and CAST. These sensitivity curves are adopted from Ref.~\cite{Aggarwal:2020olq}.
The brown curve illustrates the parameter space that can be probed by resonant cavities~\cite{Herman:2022fau}. Although the gravitational wave signals from PBH evaporation remain inaccessible to conventional interferometric detectors, future resonant cavity experiments with significantly improved sensitivity could probe the GW spectra produced by sufficiently light PBHs. 

Since the stochastic gravitational wave background behaves as an additional relativistic component in the Universe, it contributes to the effective number of neutrino species, $N_{\rm eff}${, through
\begin{equation}
\Delta N_{\rm eff} = \frac{8}{7}\left(\frac{11}{4}\right)^{4/3}\frac{\Omega_{\rm GW}h^2}{\Omega_{\gamma}h^2}\,,
\end{equation}
where $\Omega_{\rm GW}$ is the total gravitational wave energy density integrated over frequency.}
Consequently, constraints on $\Delta N_{\rm eff}$ can be translated into the upper bounds on the 
parameter space of $h_c-f$ plane. The Planck collaboration has already put a  constraint $\Delta N_{\text{eff}} < 0.30$ at 95$\%$ confidence level~\cite{Planck:2018vyg}. Future experiments, including CMB-S4 \cite{CMB-S4:2016ple}, Euclid \cite{laureijs2011euclid}, and potentially cosmic-variance-limited (CVL) CMB polarization experiments are anticipated to improve the constraint on $\Delta N_{\text{eff}}$ to $\lesssim 0.06$, $\lesssim 0.013$, and $\simeq 3.1 \times 10^{-6}$, respectively \cite{Ben-Dayan:2019gll}. These corresponding constraints are depicted with dark yellow lines in Fig. \ref{fig:GW_direct}.
The stochastic gravitational wave background originating from the direct evaporation of PBH aligns with the current $\Delta N_{\rm eff}$ constraint from Planck. Furthermore, a significant region of the parameter space could be tested by future CMB observations, particularly if next-generation measurements achieve the projected sensitivity to $\Delta N_{\rm eff}$.

\section{Conclusion}
\label{sec:conc}
We have examined non-thermal leptogenesis and Planck-scale remnant dark matter driven by primordial black hole evaporation in a minimal Pati--Salam cosmology, with the gauge symmetry broken before inflation so that the monopoles produced at the transition are diluted away. The
motivation for the embedding is that the right-handed neutrinos are then no longer optional rather they are components of the $SU(2)_R$ doublets, and their
Majorana masses are tied to the breaking scale through $M_N=y_Rv_R$. The heavy neutrino mass is therefore not a free parameter but a derived quantity, constrained by the same conditions that govern the black hole population. Requiring that the Pati--Salam symmetry be neither restored
during inflation nor regenerated by the hot plasma that follows, and that the black holes be hot enough to emit the heavy neutrinos, forces the Yukawa coupling of the Pati--Salam scalar sector well below unity and the heavy neutrino mass well below the $SU(2)_R$ breaking scale. This is a
prediction of the embedding rather than an assumption, and it is the principal way in which the gauge structure constrains what would otherwise be an unconstrained mechanism.

The baryon asymmetry can be generated in two distinct regimes depending on whether the black holes are hot enough to emit the lightest right-handed neutrino. Below roughly $10^2$ g Hawking emission populates
the heavy neutrino sector efficiently and the asymmetry is genuinely non-thermal and the requirement that it be reprocessed by the electroweak
sphalerons before they freeze out gives an upper bound on the intial mass. For heavier black holes the emission of RHNs remain Boltzmann suppressed and leptogenesis also relies on the thermal mechanism to produce the asymmetry and in this scenario the black holes evaporation also cause an entropy dilution of the asymmetry generated long before they evaporate. These two regimes respond very differently to
the possibility that quantum gravitational backreaction halts the evaporation at the Planck scale. Each black hole then leaves a single
stable remnant, and the surviving mass fraction is simply the ratio of the remnant mass to the initial mass, independent of the initial abundance. The present-day remnant fraction consequently scales as the inverse five-halves power of the initial mass, so that suppressing it demands heavier black holes rather than lighter ones. The gram-scale masses obligatory for non-thermal leptogenesis then overproduce the dark matter.  Thus Planck-scale remnants and PBH-only leptogenesis are mutually exclusive. 

However the thermal regime solves this problem. In this scenario a single initial PBH mass of order $10^6$, gives rise to the observed dark matter abundance, selects the heavy neutrino mass that reproduces the baryon asymmetry, and sets the lower limit on the Pati--Salam breaking scale. Here the heavy neutrino mass scale is light enough to be produced from the thermal plasma. The scenario therefore has the freedom to accommodate both the baryon asymmetry and the dark matter at a single point in parameter space.

The gauge sector leaves no gravitational wave signal: the pre-inflationary phase transition is inflated away and the breaking chain produces no cosmic strings. Gravitons radiated directly during evaporation peak in the ultra-high-frequency band, beyond the reach of any detector, and are constrained only through their contribution to $\Delta N_{\rm eff}$, where next-generation CMB measurements will probe a significant region of parameter space. The observationally promising channel is the tensor background sourced by the Poisson fluctuations of the black hole distribution, which for the masses selected by the remnant condition falls within the anticipated sensitivity of the future GW experiments. Since neither the baryon asymmetry nor the remnant abundance depends appreciably on the initial abundance $\beta$, this background is the only observable sensitive to it and a detection would fix this free parameter of the scenario.

\section{Acknowledgments}
S.M. acknowledges support from the IIT Goa Startup Grant [2025/SG/SM/057]. A.C. acknowledges VIT for support.

\bibliographystyle{JCAP}
\bibliography{ref}

\end{document}